\documentclass[5p,times,twocolumn,authoryear]{elsarticle}
\journal{Icarus}

\usepackage{lineno}
\usepackage{aas_macros}

\usepackage{amsmath}
\usepackage{url}

\newcommand{\revise}[1]{#1}

\usepackage[T1]{fontenc}
\begin{document}

\begin{frontmatter}



\title{Origins of Satellite-System Building Blocks I:\\
Solid Delivery During Jupiter’s Growth
}

\author[1]{Sho Shibata\corref{cor1}}
\affiliation[1]{
    organization={Department of Earth, Environmental and Planetary Sciences,  Rice University},
    addressline={MS 126},
    city={Houston},
    postcode={77005},
    state={TX},
    country={USA}
}
\ead{ss508@rice.edu}

\author[1]{Andre Izidoro}

\cortext[cor1]{Corresponding author}

\begin{abstract}
Jupiter's satellites formed within a circumplanetary disk (CPD), but the origin and delivery of the solids from which they accreted remain uncertain. Here, we investigate the delivery of planetesimal material to the CPD while consistently accounting for the growth of Jupiter and the evolution of its surrounding planetesimal reservoir. Using $N$-body simulations coupled to an evolving CPD model, we quantify the solid mass supplied through planetesimal ablation and direct capture. We find that, although proto-Jupiter rapidly opens a gap in the planetesimal disk, planetesimals continue to enter the CPD for more than $1$ Myr. Under efficient ablation, the mass deposited in the CPD can reach approximately $1 M_\oplus$, about one
order of magnitude higher than the mass delivered through direct capture. However, the low efficiency of pebble accretion within the CPD implies that forming the Galilean satellites from ablated material alone may require a very large pebble reservoir. Directly captured planetesimals, despite contributing less total mass, can instead provide satellitesimals that grow through mutual collisions and serve as seeds for subsequent pebble accretion. Our results therefore favor a hybrid formation pathway in which the Galilean satellites grow through a combination of satellitesimal and pebble accretion. We further demonstrate that even a planetesimal reservoir containing less solid material than the minimum-mass solar nebula can supply sufficient building blocks to form the Galilean satellites. The balance between the ablation and direct capture depends sensitively on the efficiency of aerodynamic heating: if ablation is weaker than commonly assumed, direct capture can become the dominant source of satellite building blocks. These results connect the formation of Jupiter and its satellites by showing how Jupiter's growth regulates both the duration and form of solid delivery to its CPD.
\end{abstract}


\begin{highlights}
\item Planetesimal delivery to Jupiter's circumplanetary disk persists for over 1 Myr.
\item Efficient ablation supplies about 1 Earth masses versus 0.1 by capture.
\item Captured planetesimals form seeds, cutting pebble demand by a factor of 5-10.
\item Required planetesimal resorvier in protosolar nebula could be less than minimum-mass solar nebula.
\item Weaker aerodynamic heating can make direct capture the dominant pathway.
\end{highlights}


\end{frontmatter}


\section{Introduction}

Jupiter and Saturn host massive systems of regular satellites. The combined mass of the Galilean satellites is approximately $0.066~M_\oplus$, corresponding to a satellite-system-to-planet mass ratio of $2.1\times10^{-4}$. Saturn's regular satellites have a total mass of approximately $0.023~M_\oplus$, yielding a similar ratio of $2.4\times10^{-4}$, although most of this mass is concentrated in Titan. Forming such massive satellite systems requires both a substantial reservoir of solid material in the circumplanetary environment and sufficiently efficient accretion onto the growing satellites. Explaining how circumplanetary disks (CPDs) acquired and retained sufficient solid material to form these massive satellite systems remains a central problem in satellite-formation theory \citep{Canup2009,Nimmo2026}.

Regular satellites are thought to have formed in gaseous disks surrounding their host planets during the final stages of giant-planet growth \citep{Lunine1982, Canup2002, Mosqueira2003a, Estrada2009}. In this framework, the satellite building blocks ultimately originated in the protoplanetary disk, but their physical form, delivery pathway, and supply history remain uncertain \citep{Nimmo2026}. Two broad delivery channels have been considered. Sufficiently small dust grains may enter the Hill sphere with the vertically accreting gas, provided that they are sufficiently stirred above the protoplanetary-disk midplane \citep{Tanigawa2014, Homma2020, Maeda2024}. 

Alternatively, heliocentric planetesimals can be captured by gas drag in the CPD and/or can deposit mass through ablation, fragmentation, or collisions. The delivery of solids by planetesimals is particularly important once proto-Jupiter exceeds the pebble-isolation mass \citep[e.g.][]{Lambrechts2014}. At this stage, the pressure maximum generated exterior to the planet's orbit strongly suppresses the inward drift of pebbles from the protoplanetary disk, limiting their direct supply to the CPD. Planetesimals can circumvent this barrier. Their interaction with the CPD can then supply solids in two ways. Ablation releases rocky and icy material into the disk, providing small solids that can contribute to satellite growth, whereas planetesimals that survive ablation and become captured may themselves serve as seeds for satellite formation.

Gas-drag capture of planetesimals by CPDs has been explored in several studies. Some models have focused on examining how capture depends on planetesimal size and encounter geometry \citep{Fujita2013, Tanigawa2014}, while other studies focused on the orbital evolution and radial distribution of captured bodies \citep{Suetsugu2016, Suetsugu2017}. Beyond the capture process itself, \citet{DAngelo2015} showed that planetesimals passing through a CPD can undergo substantial aerodynamic ablation, losing a significant fraction of their mass before either colliding with the planet, becoming permanently captured within the disk, or escaping the system. \citet{Ronnet2020} subsequently connected this ablated material to satellite formation by modeling its conversion into dust and pebbles that can be accreted by satellite seeds. However, these studies generally assume an already formed, fixed-mass giant planet and a corresponding CPD. They therefore characterize planetesimal capture and ablation under a limited set of conditions, rather than following how these processes evolve throughout Jupiter’s formation.

Because the planetesimal population reaching Jupiter’s Hill sphere is shaped by the planet’s own growth and dynamical evolution, solid delivery to the CPD cannot be treated independently of giant-planet formation. As proto-Jupiter grows, the expansion and migration of its feeding zone continually alter the reservoir of planetesimals available for accretion \citep{Zhou2007, Shiraishi2008, Shibata2019, Shibata2020, Shibata2022a, Shibata2022b, Shibata2022c}. The properties and flux of planetesimals entering the CPD, together with the amount, timing, and duration of solid delivery, should therefore evolve with the planet. Capturing this coupling requires a self-consistent model that follows Jupiter’s growth and the dynamical evolution of the surrounding planetesimal population simultaneously.

The timescale of material delivery to the CPD is particularly important for satellite formation. If Callisto's measured gravity field indeed reflects an incompletely differentiated interior, thermal-evolution calculations imply that it must have accreted over at least approximately 0.6 Myr to avoid extensive melting \citep{Barr2008}. Similarly, slow pebble-accretion models require a sustained solid flux over Myr timescales \citep{Shibaike2019}. However, a finite planetesimal reservoir within Jupiter's feeding zone can be rapidly depleted by accretion and scattering \citep{DAngelo2015,Ronnet2020}. A self-consistent model of Jupiter's growth and planetesimal dynamics is therefore essential for determining whether solid delivery to the CPD can persist for the timescales required to form the Galilean satellites.

This paper is the first in a series examining the time evolution of material supply to circumplanetary systems during the formation of their host planets. Here, we couple a Jupiter-formation model to a model of CPD evolution and determine how the capture and ablation of planetesimals evolve as Jupiter grows. A companion study will consider the concurrent formation of Jupiter and Saturn and investigate how their accretion and scattering of planetesimals in the protoplanetary disk affect the material supplied to the Jovian and Saturnian systems.

In Sec.~\ref{sec:Method}, we describe the coupled Jupiter-growth, CPD-evolution, and planetesimal-orbit model. Section~\ref{sec:Result} presents the time-dependent capture and ablation results. In Sec.~\ref{sec:satellite_formation}, we examine the implications for the formation of regular satellites. Section~\ref{sec:Discussion} discusses the model assumptions, and Sec.~\ref{sec:summary} summarizes our conclusions.

\section{Method}\label{sec:Method}

\subsection{Jupiter's formation model}
We follow the Jupiter formation model developed by \citet{Shibata2022c}, which tracks Jupiter's evolution from the onset of runaway gas accretion to the dissipation of the protoplanetary disk. During the early stage of formation, the gravitational energy released by gas accretion keeps proto-Jupiter's envelope hot and extended, with its outer radius reaching approximately the Bondi or Hill radius. The gas accretion rate is regulated by the cooling and contraction of the planetary envelope through radiative energy loss. This stage is known as the {\it attached phase}, because the planet's gaseous envelope remains smoothly connected to the surrounding protoplanetary disk.

During the attached phase, the radiative energy-loss rate increases as the planet grows \citep[e.g.][]{Ikoma2000,Tajima1997}. Once this cooling rate exceeds the maximum energy input that can be supplied by gas accreting from the surrounding protoplanetary disk, the planet transitions to the {\it detached phase}. At this point, the planetary envelope begins to contract rapidly. The proto-Jupiter's radius shrinks to approximately a few Jupiter radius within a short timescale, $\ll 10^5$~yr \citep{Valletta2021}, leaving most of the Hill sphere nearly empty. During the detached phase, gas accreting from the protoplanetary disk first enters the CPD before either accreting onto proto-Jupiter or flowing back into the protoplanetary disk.

The transition from the attached to the detached phase occurs when proto-Jupiter reaches a mass of $\sim 10$--$30\,M_\oplus$, depending on the envelope-cooling timescale. We begin our simulations at a protoplanetary disk age of $t=t_0=1~\mathrm{Myr}$, assuming that proto-Jupiter enters the detached phase at this time with a mass of $M_\mathrm{p}=10\,M_\oplus$.

Jupiter's growth is modeled self-consistently with the evolution of the protoplanetary disk. 
\revise{
The gas accretion timescale, $\tau_\mathrm{acc}$, is given by \citep{Tanigawa2002}.
\begin{align}
    \tau_\mathrm{acc} = \frac{M_\mathrm{p}}{D \Sigma_\mathrm{PPD}}, \label{eq:GasAccretionTimescale_MT}
\end{align}
where $\Sigma_\mathrm{PPD}$ is the surface density of gas in the PPD near proto-Jupiter's orbit, and $D$ is the product of the effective width of the gas accretion band and the gas inflow velocity obtained from the hydrodynamical simulations. Although $D$ increases as proto-Jupiter grows, gap opening and the feedback from gas accretion reduce $\Sigma_\mathrm{PPD}$ near proto-Jupiter's orbit. As a result, the gas accretion timescale increases as proto-Jupiter grows, regulating the evolution of the circumplanetary disk. Proto-Jupiter also migrates inward through type-II migration, with the migration rate scaled by the local gas surface density, $\Sigma_\mathrm{PPD}$ \citep{Kanagawa2018}.
}
In our model, proto-Jupiter migrates from $6.0$ to $5.2~\mathrm{au}$ and reaches Jupiter's present-day mass before the protoplanetary disk fully dissipates. The details of our model are described in \ref{app:Jupiter_model}.

\subsection{Circumplanetary disk}
The formation of a circumplanetary disk (CPD) depends on the thermal evolution of the gas within the Hill sphere \citep{Szulagyi2016}. During the detached phase, proto-Jupiter’s envelope cools via radiative losses from the photosphere and contracts on the Kelvin--Helmholtz timescale. This timescale is given by \citep{Ikoma2000, Ida2018}
\begin{align}\label{eq:Kelvin-Helmholtz_time}
    \tau_\mathrm{KH} \sim 10^3 \mathrm{yr} \left( \frac{M_\mathrm{p}}{10 M_\oplus} \right)^{-3} \left( \frac{\kappa_\mathrm{env}}{0.01 \mathrm{cm^2/g}} \right),
\end{align}
where $M_\mathrm{p}$ is the planetary mass, and $\kappa_\mathrm{env}$ is the opacity in the planetary envelope. The envelope opacity depends on the size and abundance of dust grains in the planetary envelope. Grain-growth models suggest that $\kappa_\mathrm{env}$ can be reduced by several orders of magnitude relative to the interstellar-medium opacity \citep{Mordasini2014,Ormel2014}. For $\kappa_\mathrm{env}\leq0.01~\mathrm{cm^2,g^{-1}}$, the planetary envelope contracts rapidly, allowing a compact CPD to form within $\sim10^3~\mathrm{yr}$. Based on this picture of rapid contraction, we assume that the CPD has already formed at the beginning of the simulations. Proto-Jupiter's photospheric radius contracts to a few Jupiter radii on the Kelvin--Helmholtz timescale and subsequently evolves only slowly over gigayear timescales \citep{Valletta2021}. We therefore set the radius of proto-Jupiter, $R_\mathrm{p}$, to $1.6~R_\mathrm{Jup}$, where $R_\mathrm{Jup}$ denotes Jupiter's present-day radius. The possible effects of higher envelope opacity and slower envelope's contraction are discussed later.

To model the evolution of the CPD surface density, we adopt the actively supplied accretion-disk model of \citet{Canup2002, Canup2006}. In this model, gas is continuously delivered from the protoplanetary disk to the CPD. The resulting surface density, $\Sigma_\mathrm{CPD}$, is given by
\begin{align}\label{eq:surface_density_CPD}
    \Sigma_\mathrm{CPD} = \dfrac{4F_\mathrm{gas}}{15\pi\nu}
    \begin{cases}
        \left[
            \dfrac{5}{4}
            -\sqrt{\dfrac{r_\mathrm{c}}{r_\mathrm{d}}}
            -\dfrac{1}{4}\left(\dfrac{d}{r_\mathrm{c}}\right)^2
        \right],
        & d \le r_\mathrm{c}, \\[1.2ex]
        \left[
            \sqrt{\dfrac{r_\mathrm{c}}{d}}
            -\sqrt{\dfrac{r_\mathrm{c}}{r_\mathrm{d}}}
        \right],
        & d > r_\mathrm{c}.
    \end{cases}
\end{align}
where $d$ is the radial distance from proto-Jupiter, $r_\mathrm{c}$ is the characteristic radius within which gas is supplied to the disk, $r_\mathrm{d}$ is the outer radius of the viscously spreading disk, and $F_\mathrm{gas}$ is the total gas infall rate onto the disk within $d<r_\mathrm{c}$. Following previous models \citep{Sasaki2010, Ronnet2020}, we set $r_\mathrm{c}=30~R_\mathrm{Jup}$ and $r_\mathrm{d}=0.2~R_\mathrm{H}$, where $R_\mathrm{H}$ is the Hill radius. 
\revise{
As proto-Jupiter grows, $r_\mathrm{d}$ increases from $\sim60~R_\mathrm{Jup}$ to $\sim150~R_\mathrm{Jup}$.
}
These characteristic radii are motivated by three-dimensional hydrodynamical simulations \citep{Machida2008, Machida2009, Tanigawa2012}. We set the infall rate to $F_\mathrm{gas}=M_\mathrm{p}/\tau_\mathrm{acc}$, where $M_\mathrm{p}$ is the mass of proto-Jupiter and $\tau_\mathrm{acc}$ is its gas-accretion timescale, given by the Jupiter's formation model (see \ref{app:Jupiter_model}). Here, $\nu$ is the kinematic viscosity of the CPD. We adopt the $\alpha$-viscosity prescription \citep{Shakura+1973} and set the viscosity parameter to $\alpha_\mathrm{CPD}=10^{-3}$. 
\revise{
In our fiducial model, we neglect the dust delivery entrained in the accreting H/He gas to the CPD.
}

The temperature structure of a viscously evolving disk is controlled by viscous heating and irradiation. Recent disk-wind models suggest that viscous heating may be weaker than in classical viscously evolving disks, because angular momentum can be extracted by winds launched from the disk surface \citep{Bai2017, Mori2019}. A similar argument may also apply to CPDs \citep{Fujii2014, Shibaike2023}. Motivated by this disk-wind picture, we adopt an irradiation-dominated CPD as our reference model. 

In the irradiation dominated disk, the gas temperature is separeted into two regimes \citep{Ruden1991, Ronnet2020}. In the inner disk region, where the photospheric height, $H_\mathrm{ph}$, is smaller than the planetary radius, the temperature profile is given by \citep{Ronnet2020}
\begin{align}
    T_\mathrm{thin} = 190 \mathrm{K} \left( \frac{M_\mathrm{p}}{M_\mathrm{Jup}} \right)^{1/2} \left( \frac{\tau_\mathrm{acc}}{5 \mathrm{Myr}} \right)^{-1/4} \left( \frac{d}{10 R_\mathrm{Jup}} \right)^{-3/4},
\end{align}
where $M_\mathrm{Jup}$ denotes Jupiter's present-day mass. In the outer disk region, where the grazing angle changes with the disk's flare,  the temperature profile follows \citep{Ronnet2020}
\begin{align}
    T_\mathrm{flar} =& 190 \mathrm{K} \left( \frac{M_\mathrm{p}}{M_\mathrm{Jup}} \right)^{3/7} \left( \frac{\tau_\mathrm{acc}}{5 \mathrm{Myr}} \right)^{-2/7} \nonumber\\ \times&\left( \frac{d}{10 R_\mathrm{Jup}} \right)^{-3/7}
    \left( \frac{R_\mathrm{p}}{1.6 R_\mathrm{Jup}} \right)^{-2/7} \left( \frac{\chi}{4} \right)^{2/7},
\end{align}
where $\chi$ is defined as $\chi = H_\mathrm{ph}/H_\mathrm{CPD}$, and $H_\mathrm{CPD}$ is the scale height of CPD. $\chi$ is of order of unity, and depends on the opacity of the disk. In this study, we assume $\chi=4$. We calculate the temperature of CPD, $T_\mathrm{CPD}$, as
\begin{align}
    T_\mathrm{CPD} = \max \left( T_\mathrm{thin}, T_\mathrm{flare} \right).
\end{align}

\begin{figure}
    \centering
    \includegraphics[width=0.9\linewidth]{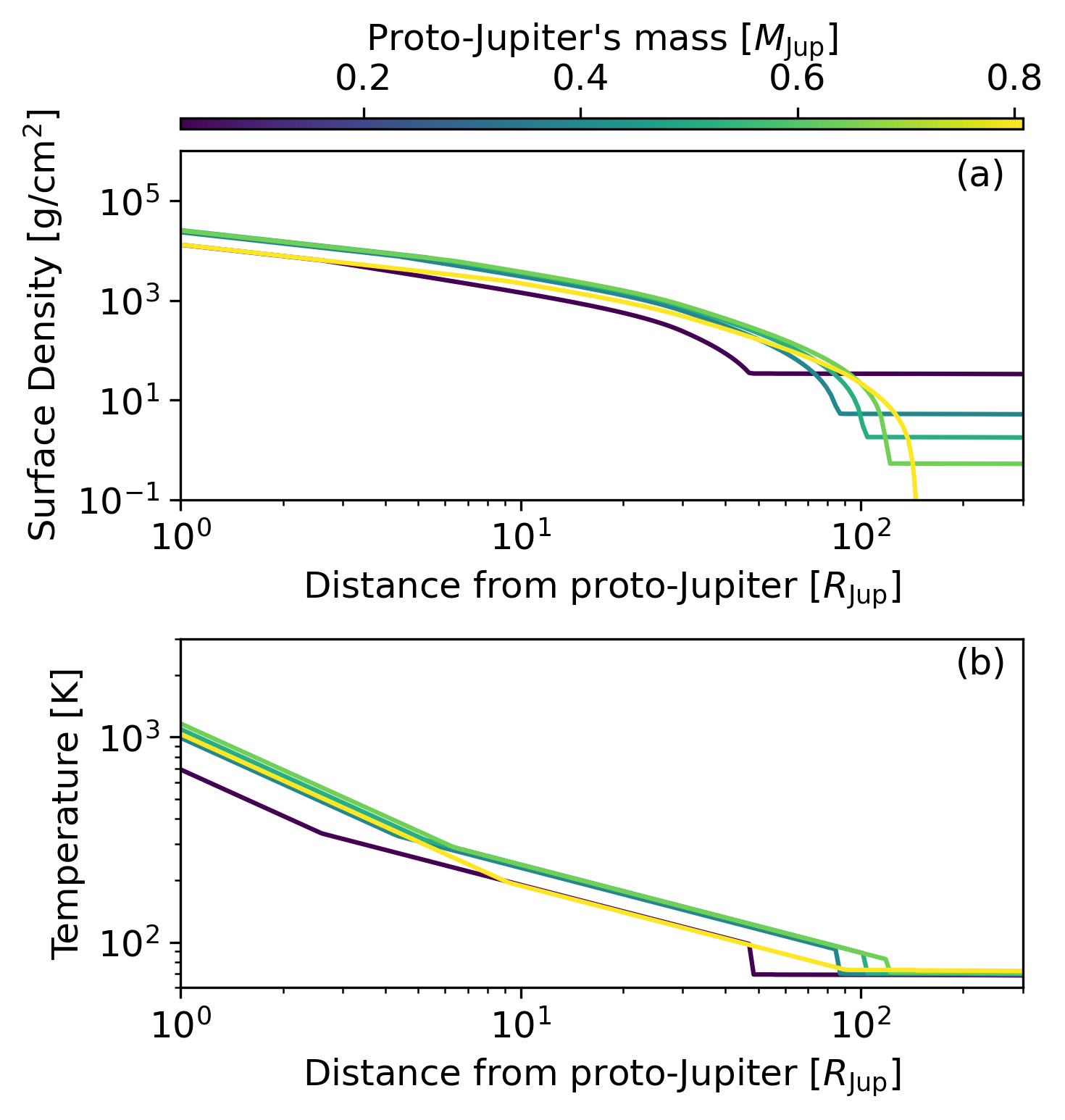}
    \caption{
    Circumplanetary disk (CPD) model used in this study.
    {\bf (a):} Gas surface density profile as a function of distance from proto-Jupiter. Different colors indicate different proto-Jupiter masses. 
    {\bf (b):} Temperature profile as a function of distance from proto-Jupiter. Each line corresponds to a different phase of proto-Jupiter's growth, as indicated in the color bar at the top of the panel.
    \revise{
    The CPD is assumed to connect to the protoplanetary disk (PPD) at the bottom of the gap opened by proto-Jupiter, where the surface densities are matched ($\Sigma_\mathrm{CPD}=\Sigma_\mathrm{PPD}$). The flat portions of the profiles in both panels correspond to the PPD.
    }
    }
    \label{fig:CPD}
\end{figure}

Figure~\ref{fig:CPD} shows the surface density and temperature profiles of the CPD. The CPD's outer edge, $r_\mathrm{d}$, is connected to the gap bottom of the protoplanetary disk. Because both the planetary mass and gas accretion rate evolve with time, the irradiation heating from proto-Jupiter also changes over the Jupiter's formation. We summarise the parameters used in our CPD disk model in Tab.~\ref{tb:model_disks}.

\begin{table}
	\centering
	\begin{tabular}{ llllll } 
		\hline
		\hline
		$r_\mathrm{c}$ 		& Critical radius for the infall gas & $30 R_\mathrm{Jup}$ &	\\
		$r_\mathrm{d}$ 		& Diffused-out disk outer edge & $0.2 R_\mathrm{H}$ &	\\
		$\chi$ 		& Scaling parameter of photospheric height & $4$ &	\\
		$\alpha_\mathrm{CPD}$ 		& $\alpha$-viscosity parameter in CPD   & $10^{-3}$ &	\\
		\hline
	\end{tabular}
	\caption{
    Parameters used in the CPD model.
    }
    \label{tb:model_disks}
\end{table}

\subsection{Disk gas drag and ablation}

Planetesimals moving in gaseous disk feels aerodynamic gas drag. The accelaration through the gas drag is given by \citep{Adachi1976}
\begin{align}\label{eq:reduced_gas_drag}
  {\bf a}_\mathrm{gas} = -\frac{{\bf u}}{\tau_\mathrm{aero}}, 
\end{align}
with
\begin{align}\label{eq:damping_timescale}
  \tau_\mathrm{aero} = \frac{8 \rho_\mathrm{s} R_\mathrm{s}}{3 C_\mathrm{d} \rho_\mathrm{g} u},
\end{align}
where ${\bf u}$ is the velocity relative to the ambient gas, $\tau_\mathrm{aero}$ is the damping timescale of aerodynamic gas drag, $C_\mathrm{d}$ is the non-dimensional drag coefficient (see \ref{app:Cd}), $\rho_\mathrm{g}$ is the density of ambient gas, $\rho_\mathrm{s}$ is the bulk density of planetesimals, and $R_\mathrm{s}$ is the planetesimal's radius. The density and velocity of the ambient gas are calculated from the PPD and CPD models. Both disks are assumed to be vertically isothermal. 
\revise{
The CPD and PPD are connected at the radius where their gas surface densities match ($\Sigma_\mathrm{CPD}=\Sigma_\mathrm{PPD}$). Gas is assumed to orbit proto-Jupiter within the CPD, whereas it orbits the central star within the PPD.
}
In both regions, the gas moves at a sub-Keplerian azimuthal velocity owing to radial pressure support.

Aerodynamic gas drag damps the orbits of planetesimals and also heats their surfaces. The evolution of the planetesimal's surface temperature, $T_\mathrm{s}$, is described by the energy balance equation \citep{Popova2006, DAngelo2015, Ribeiro2024}:
\begin{align}\label{eq:T_surf_evolution}
    M_\mathrm{s} C_\mathrm{s} \frac{\mathrm{d} T_\mathrm{s}}{\mathrm{d} t} =& \frac{\pi}{2} C_\mathrm{h} \rho_\mathrm{g} R_\mathrm{s}^2 u^3 \nonumber\\
    &+ 4 \pi R_\mathrm{s}^2 \epsilon_\mathrm{s} \sigma_\mathrm{SB} \left(T_\mathrm{g}^4 - T_\mathrm{s}^4 \right) \nonumber\\
    &+ L_\mathrm{s} \frac{\mathrm{d} M_\mathrm{s}}{\mathrm{d} t},
\end{align}
where $M_\mathrm{s}$ is the planetesimal mass, $C_\mathrm{s}$ is the specific heat of the planetesimal, $C_\mathrm{h}$ is the heat transfer coefficient, $\epsilon_\mathrm{s}$ is the thermal emissivity of the planetesimal, $\sigma_\mathrm{SB}$ is the Stefan--Boltzmann constant, $T_\mathrm{g}$ is the ambient gas temperature, and $L_\mathrm{s}$ is the latent heat of the ablating material. We set $\epsilon_\mathrm{s}=1$ in this work. The heat transfer coefficient $C_\mathrm{h}$ represents the fraction of the collisional kinetic energy transferred to the planetesimal as heat. Following \citet{DAngelo2015}, we adopt $C_\mathrm{h}=C_\mathrm{d}/4$, which is expected to provide an upper limit on the heating efficiency, in our reference models. We also perform a parameter study in which $C_\mathrm{h}$ is varied in Sec.~\ref{sec:discussion_Ch}.

The ablation rate of evaporated material is given by \citep{DAngelo2015}
\begin{align}\label{eq:ablation_rate}
    \frac{\mathrm{d} M_\mathrm{s}}{\mathrm{d} t}
    &= -\frac{1}{4} \rho_\mathrm{evap} \bar{v}_\mathrm{th} A_\mathrm{s}, \nonumber \\
    &= -\frac{1}{4}
    \frac{\mu_\mathrm{s} m_\mathrm{H} P_\mathrm{v}}{k_\mathrm{B} T_\mathrm{s}}
    \sqrt{\frac{8}{\pi} \frac{k_\mathrm{B}T_\mathrm{s}}{\mu_\mathrm{s} m_\mathrm{H}}}
    A_\mathrm{s}, \nonumber \\
    &= -4\pi R_\mathrm{s}^2 P_\mathrm{v} (T_\mathrm{s})
    \sqrt{\frac{\mu_\mathrm{s} m_\mathrm{H}}{2\pi k_\mathrm{B} T_\mathrm{s}}},
\end{align}
where $\rho_\mathrm{evap}$ is the density of the evaporated gas, $\bar{v}_\mathrm{th}$ is its mean thermal velocity, and $A_\mathrm{s}=4\pi R_\mathrm{s}^2$ is the surface area of the planetesimal, $P_\mathrm{v}$ is the vapor pressure of the material, $\mu_\mathrm{s}$ is the mean molecular weight of the material, $m_\mathrm{H}$ is the mass of hydrogen atom, and $k_\mathrm{B}$ is Boltzmann constant. 

\citet{Ronnet2020} shows that the heating and cooling timescales of planetesimal surface is much shorter than the orbital timescale. By assuming the equilibrium condition in Eq.~\ref{eq:T_surf_evolution}, we get
\begin{align}\label{eq:T_surface}
    {T_\mathrm{s}}^{4} = {T_\mathrm{g}}^{4} 
        + \frac{C_\mathrm{h} \rho_\mathrm{g} u^3}{8 \epsilon_\mathrm{s} \sigma_\mathrm{SB}}
        - \frac{P_\mathrm{v}}{\epsilon_\mathrm{s} \sigma_\mathrm{SB}} \sqrt{\frac{\mu_\mathrm{s} m_\mathrm{H}}{2\pi k_\mathrm{B}T_\mathrm{s}}} L_\mathrm{s}.
\end{align}

At each timestep, we solve Eqs.~\ref{eq:ablation_rate} and \ref{eq:T_surface} for each planetesimal using the bisection method. We set the planetesimal composition as undifferetiated ice-rock mixture with 6:4 mass fraction. 
\revise{
We assume that the rocky grains embedded in the ice are also released and entrained in the surrounding gas flow as the ice sublimates, rather than remaining on the planetesimal surface to form a refractory crust.
}
The parameters related to the planetesimal composition are extracted from \citet{DAngelo2015} and summarised in Tab.~\ref{tb:model_ablation}. The vapor-pressure prescription, $P_\mathrm{v}$, follows that of \citet{DAngelo2015}; see \ref{app:vaper_pressure} for details.
\revise{
Throughout this paper, we neglect the feedback of gas released by planetesimal ablation on the disk structures.
}

\begin{table}
	\centering
	\begin{tabular}{ llllll } 
		\hline
		\hline
        $f_\mathrm{abl}$ & mass fraction of ablating material & 0.6 \\
        $\rho_\mathrm{s}$ 		& bulk density & 1.33 g/cm$^3$ \\
        $\mu_\mathrm{s}$ 		& mean molecular weight & 25.0 \\
        $L_\mathrm{s}$ 		& latent heat & $2.83\times10^{10}$ erg/g \\
        \hline
	\end{tabular}
	\caption{
	Parameters used in ablation model.
    }
    \label{tb:model_ablation}
\end{table}

\subsection{Numerical setup}

Our N-body simulations are performed using the REBOUND code \citep{Rein2012} with the IAS-15 integrator. We integrate orbits up to $5\times10^6$ yr.

Planetesimals are treated as super-particles, and their mutual gravitational interactions are neglected. In our fiducial simulations, we distribute 3000 particles between 3.78 and 7.42 au with uniform orbital spacing, covering the feeding zone of a Jupiter-mass planet. The mass represented by each super-particle is given by
\begin{align}
    m_\mathrm{sp} \left(a_0\right) = 2 \pi a_0 \Sigma_\mathrm{plts} \left( a_0 \right) \Delta a_0,
\end{align}
where $\Sigma_\mathrm{plts}$ is the surface density of planetesimals, $a_0$ is the initial semi-major axis of a super-particle, and $\Delta a_0$ is the orbital separation to the adjuscent super-particle, which is 0.00121 au in our fiducial simulations. We adopt the following planetesimal surface density profile:
\begin{align}
    \Sigma_\mathrm{plts} = f_\mathrm{PPD}\Sigma_{\mathrm{plts},0} \left(\frac{r}{5.2~\mathrm{au}}\right)^{-1},
\end{align}
where $f_\mathrm{PPD}$ is a scaling factor, $r$ is the radial distance from the central star, and $\Sigma_{\mathrm{plts},0}$ is set to $2.53~\mathrm{g\,/cm^{2}}$, corresponding to the planetesimal surface density of the minimum-mass solar nebula at 5.2 au \citep{Hayashi1981}. Unless otherwise noted, we adopt $f_\mathrm{PPD}=1$. This setup has $10.7 M_\oplus$ of planetesimals in total, and each super-particle represents a mass of $m_{\rm sp}=3.56\times10^{-3} M_\oplus$. The initial eccentricities and inclinations of planetesimals are drawn from Rayleigh distributions with rms values satisfying $\langle e^2\rangle^{1/2}=2\langle i^2\rangle^{1/2}=10^{-3}$. The remaining orbital elements are uniformly selected between 0 and 360 degrees.

The initial physical size of the super-particles, $R_\mathrm{s}$, which determines the strength of gas drag and ablation, is treated as a free parameter. We vary $R_\mathrm{s}$ from $0.1$ to $100~\mathrm{km}$, covering the plausible range of planetesimal sizes around proto-Jupiter when fragmentation is taken into account.

We record the mass and location of ablated materials as they evolve. Planetesimals are removed from the simulation when they satisfy one of following three conditions: collision, evaporation, or capture. A collision is considered to occur when the relative distance between a planetesimal and proto-Jupiter becomes smaller than the physical radius of proto-Jupiter. We classify a planetesimal as evaporated when its radius decreases below 10 m due to ablation. Finally, we classify a planetesimal as captured by the CPD if it remains inside the CPD for more than 100 orbits around proto-Jupiter. To confirm that the planetesimals can remain there without significant evaporation, we also require that their eccentricities, calculated with respect to proto-Jupiter, decrease below 0.1. At such low eccentricities, the relative velocity with respect to the CPD gas is reduced, making aerodynamic heating much less efficient.

\section{Result} \label{sec:Result}

\subsection{Orbital evolution of planetesimals} 

\begin{figure}
    \centering
    \includegraphics[width=0.9\linewidth]{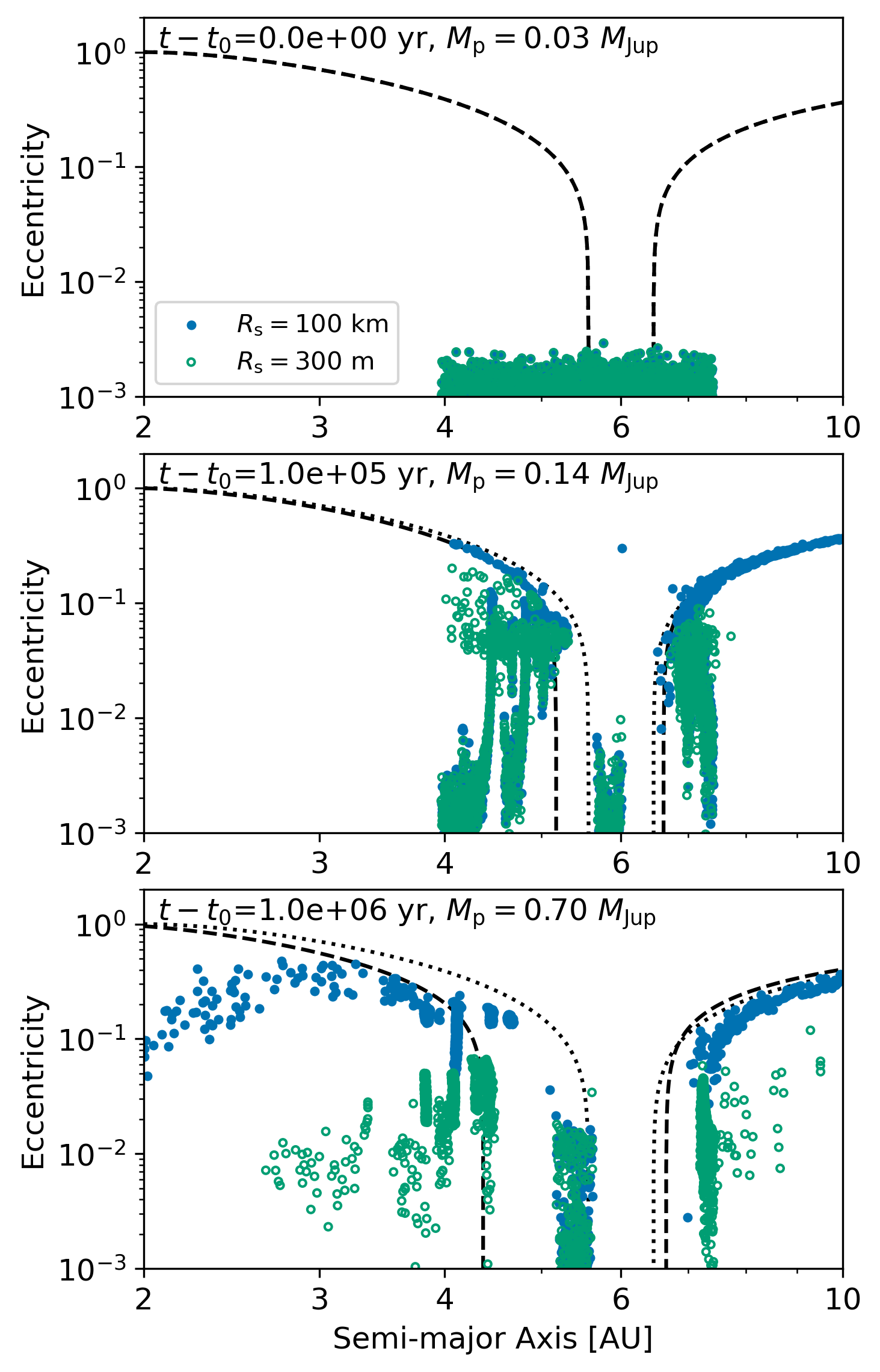}
    \caption{
        Snapshots from our numerical simulations showing the semi-major axis and eccentricity of planetesimals. The blue and green points correspond to the simulations with $R_\mathrm{s}=100$ km and $300$ m, respectively. Time progresses from the top panels to the bottom panels, as annotated in the upper left of each panel. The black dotted line indicates the boundary of the feeding zone at $t-t_0=0$ yr. The feeding zone shifts and expands as the planet migrates and grows in mass. The black dashed line shows the boundary of the feeding zone at each time. We assume $\cos i \simeq 1$ for ploting the feeding zone boundaries. 
    }
    \label{fig:orbs}
\end{figure}

Orbits of planetesimals around a growing proto-Jupiter evolve through gravitational scattering by the proto-Jupiter and aerodynamic gas drag from the disks. Figure~\ref{fig:orbs} shows snapshots of our $N$-body simulations with planetesimal sizes corresponding to $R_\mathrm{s}=100~\mathrm{km}$ (blue points) and $300~\mathrm{m}$ (green points). We also plot the boundary of the feeding zone, defined as the region with positive Jacobi energy \citep[e.g.,][]{Hayashi1977},
\begin{align}
    E_\mathrm{J} = -\frac{a_\mathrm{p}}{2a} -\sqrt{a\left(1-e^2\right)} \cos i +\frac{3}{2} +\frac{9}{2} h^2 + O \left( h^3 \right),
\end{align}
where $h=R_\mathrm{H}/a_\mathrm{p}$ is the reduced Hill radius, where $R_\mathrm{H}$ and $a_\mathrm{p}$ are the Hill radius and semimajor axis of proto-Jupiter, respectively, and $a$, $e$, and $i$ are the semimajor axis, eccentricity, and inclination of a planetesimal. At the beginning of the simulation, the feeding zone is narrow, and only a relatively small number of planetesimals are located within it. As proto-Jupiter grows, the feeding zone expands and engulfs additional planetesimals. In addition, planetary migration shifts the feeding zone, causing planetesimals to enter it \citep[e.g.][]{Tanaka1999, Zhou2007, Shiraishi2008, Shibata2019, Shibata2020, Shibata2022b}.

Gravitational scattering by proto-Jupiter excites the eccentricities of planetesimals. Planetesimals with higher eccentricities experience stronger aerodynamic drag from the surrounding disk gas, resulting in eccentricity damping and radial drift: which tend to remove them from the feeding zone. Because gas drag is more efficient for smaller planetesimals, planetesimals of 300 m tend to maintain lower eccentricities and are removed from the feeding zone more rapidly (see green dots in Figure \ref{fig:orbs}). Planetesimals of 100~km, however, tend to \revise{stay} in the feeding zone for longer, increasing their chances of entering the CPD \citep{Shibata2019, Shibata2022c}.

\begin{figure}
    \centering
    \includegraphics[width=0.9\linewidth]{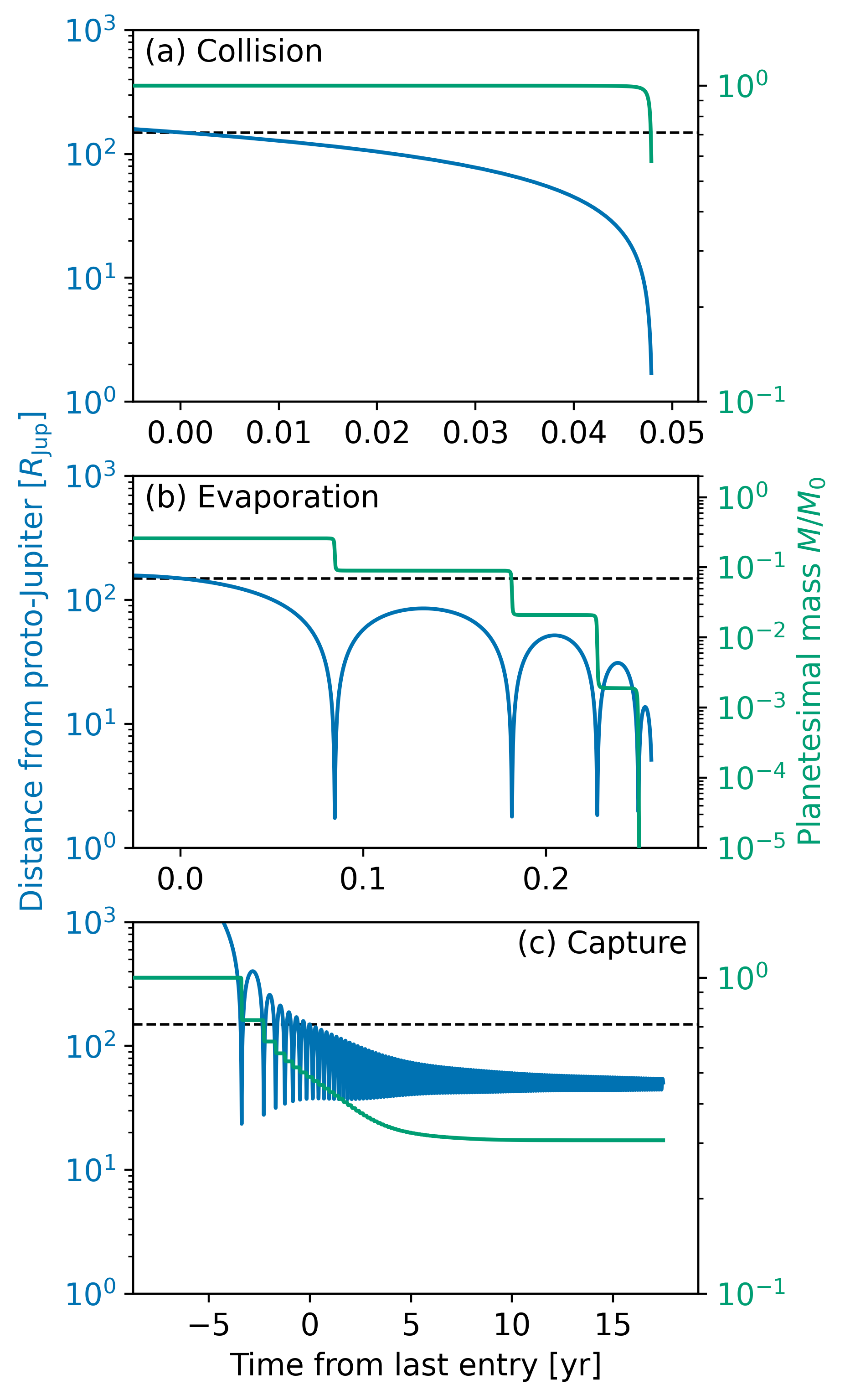}
    \caption{
    Evolution of planetesimals that ultimately undergo collision, evaporation, and capture, shown from top to bottom. The blue lines show the distance from proto-Jupiter, with values indicated on the left vertical axis, while the green lines show the planetesimal mass. The horizontal dotted line marks the outer edge of the CPD. Time is measured from the planetesimal's final entry into the CPD.
    }
    \label{fig:Fates}
\end{figure}

Figure~\ref{fig:Fates} shows the time evolution of three representative planetesimals that enter the Hill sphere of proto-Jupiter and subsequently interact with the CPD. The planetesimals shown in this figure represent three different outcomes: collision, evaporation, and capture. In panel~(a), the planetesimal directly collides with proto-Jupiter. Before the collision, it experiences strong gas drag from the CPD and loses approximately $40\%$ of its mass through ablation. In panels~(b) and (c), the planetesimals are gravitationally trapped by proto-Jupiter. The planetesimal in panel~(b) completely evaporates through ablation, whereas the planetesimal in panel~(c) completes orbital circularization before it fully evaporates and settles onto a stable orbit within the CPD. In this case, the captured planetesimal orbits around $50$ Jupiter's radius, and its mass is reduced to 30 \% of its initial mass. 

The main factor that distinguishes evaporation from capture is the gas temperature at the location where the planetesimal is trapped. In panel~(b), the pericenter of the planetesimal is smaller than $10 R_\mathrm{Jup}$, where the disk temperature exceeds the water sublimation temperature. Therefore, to be captured by the CPD without complete evaporation, planetesimals need to be trapped in the outer, cooler regions of the CPD.

The ablation of planetesimals releases rocky and icy material into the CPD, providing the raw material for the growth of the Jovian satellites. Planetesimals that survive ablation and become captured within the CPD may instead act as seeds for satellite formation, subsequently growing through the accretion of smaller planetesimals (satellitesimals) and/or pebbles.

\subsection{Case of large planetesimals with $100~\mathrm{km}$ radii}

\subsubsection{Material Supply to the CPD}

\begin{figure}
    \centering
    \includegraphics[width=1.0\linewidth]{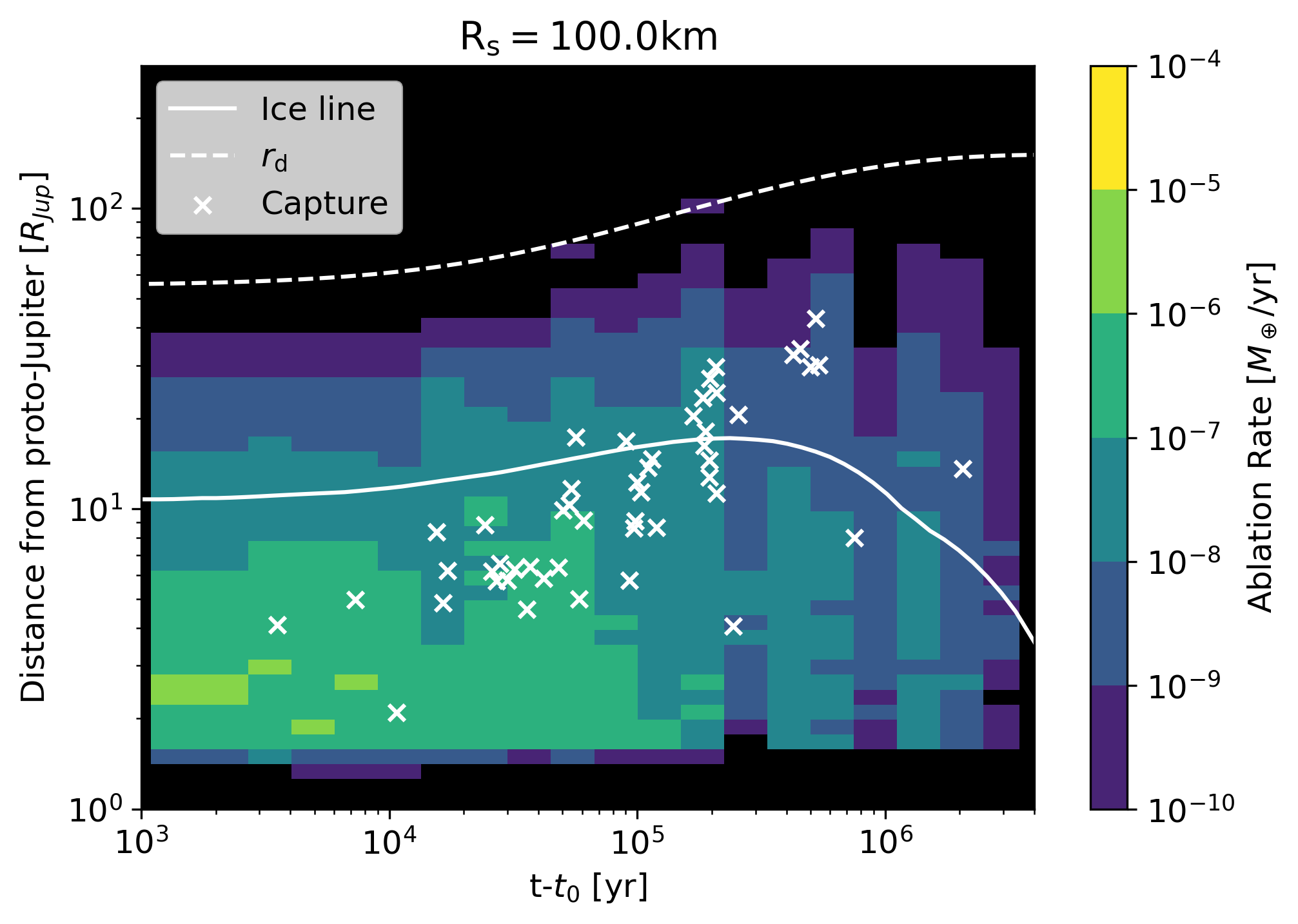}
    \caption{
    Map of the planetesimal ablation rate in the time--distance plane. Here, we show the case with $R_\mathrm{s}=100~\mathrm{km}$. The solid and dashed lines indicate the water ice line and the outer edge of the CPD, $r_\mathrm{d}$, respectively. Regions where the ablation rate is lower than $10^{-10}~M_\oplus~\mathrm{yr}^{-1}$ are shown in black. The crosses mark the locations at which planetesimals are captured. 51 super-particles are captured in this case.
    }
    \label{fig:t_r_ablation}
\end{figure}

First, we examine the ablation and capture of planetesimals with $R_\mathrm{s}=100~\mathrm{km}$. Figure~\ref{fig:t_r_ablation} provides an overview of material delivery to the CPD. The solid line indicates the location of the water ice line. As proto-Jupiter grows, the CPD initially becomes hotter because the planet's gravitational potential deepens and the energy released by gas accretion increases. At later times, however, the CPD begins to cool as the gas accretion rate declines owing to the gaseous gap opening around proto-Jupiter and the dispersal of the protoplanetary disk. Consequently, the water ice line first moves outward and then migrates inward. At the same time, the CPD expands as proto-Jupiter grows because its Hill sphere becomes larger. Accordingly, the outer edge of the disk, shown by the dashed line, moves outward with time.

The ablation rate is shown by the color map. As the CPD evolves, the region in which ablation occurs extends to larger distances from proto-Jupiter. The crosses mark the locations where planetesimals are captured. The capture region also shifts outward as the CPD expands.

Both the ablated material and the captured planetesimals provide building blocks for satellite formation. In the following sections, we examine how ablation and capture proceed during the growth of proto-Jupiter.

\subsubsection{Time evolution of the planetesimal ablation process}

\begin{figure}
    \centering
    \includegraphics[width=0.9\linewidth]{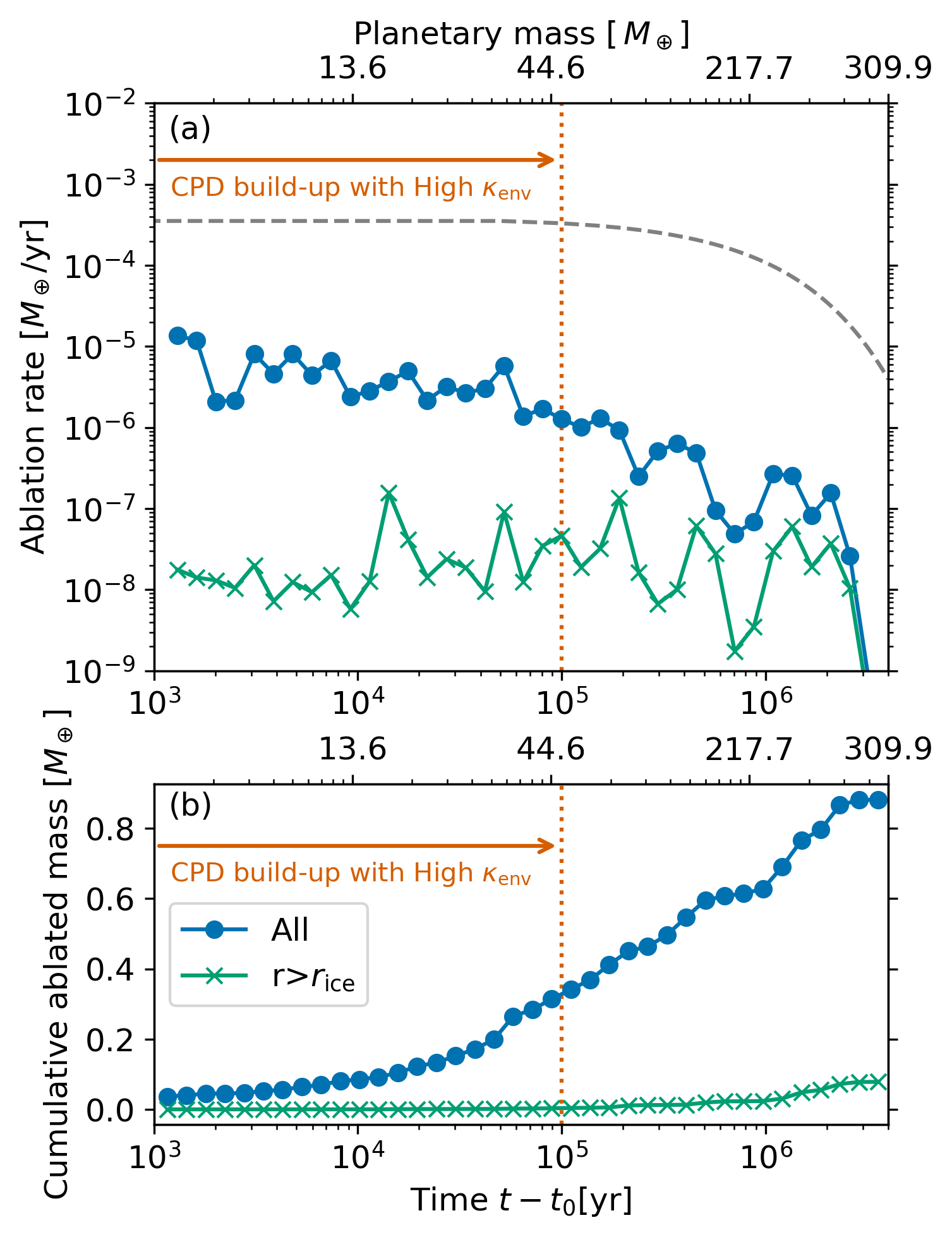}
    \caption{
        Time evolution of the ablation for the case with 100 km planetesimals. 
        {\bf (a)} Time evolution of the ablation rate. The blue and green lines show the ablation rates in the entire CPD and outside the water ice line in the CPD, respectively. The gray line is the gas accretion rate onto proto-Jupiter. The vertical dotted lines correspond to the envelope's contraction timescales for high opacity ($\kappa_\mathrm{env}=1.0\ \mathrm{cm^2/g}$), which can be interpreted as the CPD formation timescale.
        {\bf (b)} Time evolution of the cumulative ablated mass. The plots and lines are same as panel-(a). 
    }
    \label{fig:time_dMabl_100km}
\end{figure}

Figure~\ref{fig:time_dMabl_100km} shows the time evolution of planetesimal ablation. In panel~(a), we plot the ablation rate of planetesimals as a function of simulation time. The blue symbols show the total ablation rate integrated over the entire CPD. The ablation rate is as high as $10^{-5}~M_\oplus~\mathrm{yr}^{-1}$ in the early stage and decreases with time. This decrease is mainly caused by gap formation in the planetesimal disk, as shown in Fig.~\ref{fig:orbs}. Once the gap opens, the flux of planetesimals entering the feeding zone is controlled by the growth and migration timescales of proto-Jupiter. Since both timescales become longer as proto-Jupiter grows, the planetesimal ablation rate continues to decrease.

We recall that in our nominal simulations we adopted a rapid CPD build-up by assuming a low envelope's opacity of $\kappa_\mathrm{env}=0.01 \mathrm{cm^2/g}$. If the envelope in reality cools more slowly because of a higher opacity, such as $\kappa_\mathrm{env}=1.0~\mathrm{cm^2/g}$,
the CPD build-up may take up to $10^{5}~\mathrm{yr}$, as indicated by the vertical dotted line in Fig.~\ref{fig:time_dMabl_100km}. In this case, material ablated from planetesimals during $t-t_0<10^5~\mathrm{yr}$ would be trapped within the extended envelope of proto-Jupiter and might subsequently be accreted onto the planet as the envelope contracts. Therefore, only material ablated after this contraction phase would be probably available as building blocks for satellite formation.

Panel~(b) shows the cumulative ablated mass of planetesimals. Although the ablation rate decreases with time, ablation continues throughout Jupiter's formation. The total ablated mass is $0.88~M_\oplus$. If only material ablated after $t-t_0>10^5~\mathrm{yr}$ is considered, corresponding to a scenario in which CPD formation is delayed by the slower contraction of a high-opacity envelope, the available mass is reduced to $0.57~M_\oplus$. Note that these masses scale linearly with the mass of the planetesimal reservoir in the protoplanetary disk, parameterized by $f_\mathrm{PPD}$. 

For Jovian satellite formation, the location where the ablation proceeds is important because Ganymede and Callisto contain large amounts of water and therefore must have formed by accreting icy building blocks beyond the water ice line. The green crosses show the ablation rate contributed by planetesimals that ablate outside the water ice line. Comparing these values with the blue symbols, we find that most ablation occurs inside the water ice line during the early formation stage, $\lesssim 10^6~\mathrm{yr}$. Ablation outside the water ice line becomes effective only in the late formation stage, $\gtrsim10^6~\mathrm{yr}$. In the reference case, a total of $0.88~M_\oplus$ of planetesimal material is ablated within the CPD, of which $0.078~M_\oplus$ is ablated outside the water ice line.

\revise{
Finally, we discuss a potential uncertainty associated with the assumed initial distribution of planetesimals. In our simulations, we initially distribute planetesimals throughout the feeding zone of proto-Jupiter. However, a gap in the planetesimal disk may have already formed when proto-Jupiter reaches its initial mass of $0.03~M_\mathrm{Jup}$. We find that planetesimals initially located within the feeding zone are rapidly depleted within $\sim10^4~\mathrm{yr}$, consistent with the $N$-body simulations of \citet{Ronnet2020}. After this initial phase, planetesimals entering the Jovian system originate primarily from regions initially outside the feeding zone. Indeed, additional analysis with an initially depleted feeding zone show no significant difference in the rate of material delivery to the CPD after $3\times10^4~\mathrm{yr}$. Therefore, we conclude that the assumed initial planetesimal distribution has little effect on the long-term delivery of solid material to the CPD.
}

\subsubsection{Satelitesimal ring from captured planetesimals}

\begin{figure}
    \centering
    \includegraphics[width=0.9\linewidth]{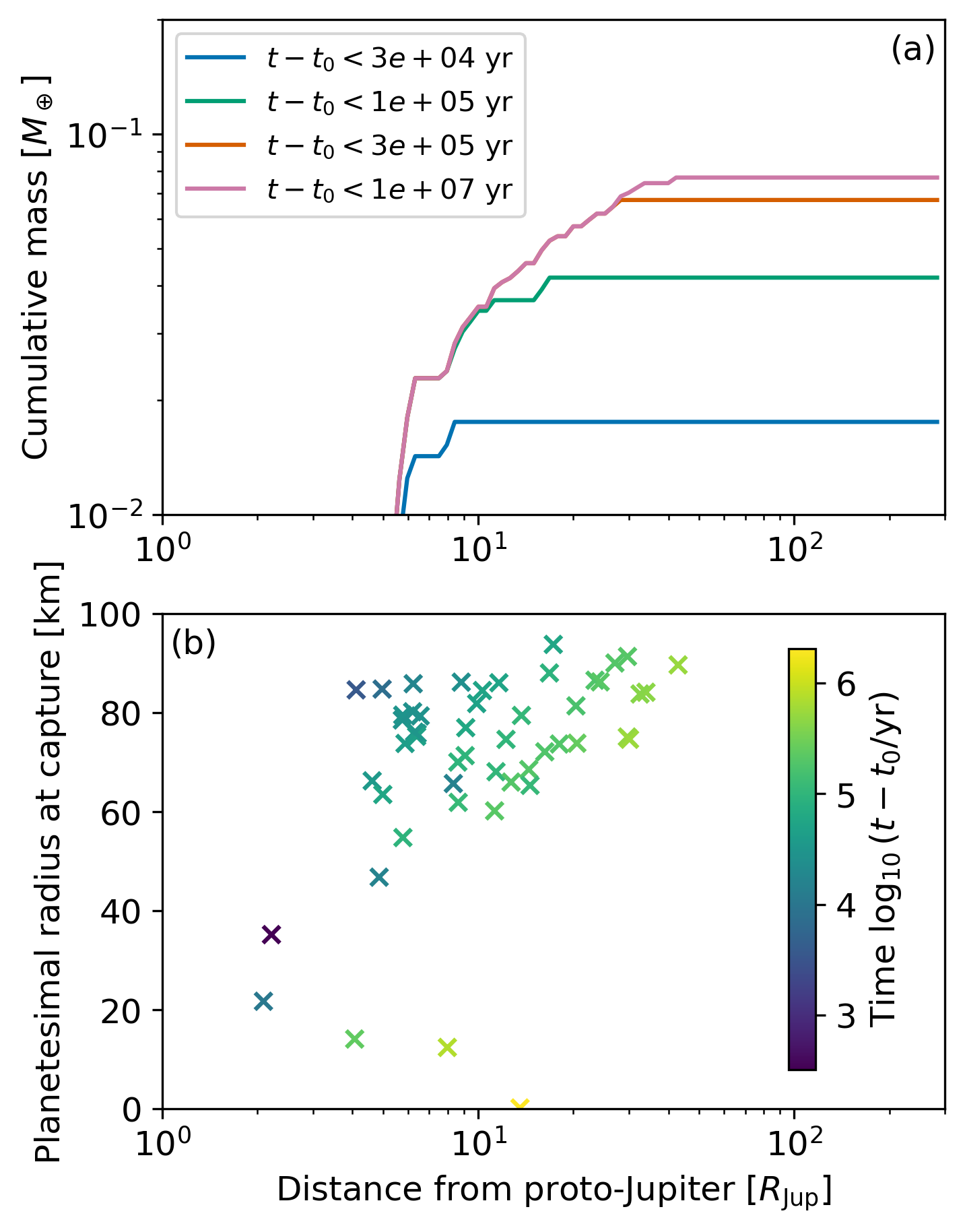}
    \caption{
        {\bf (a)} Cumulative mass distribution of captured planetesimals as a function of the distance from proto-Jupiter for planetesimals with $R_\mathrm{s}=100$ km. Different colors correspond to different time intervals during proto-Jupiter's formation as indicated in the legend box. 
        {\bf (b)} The radius of captured planetesimals v.s. the distance from proto-Jupiter. The color corresponds to the time when the planetesimal is captured by the CPD.
    }
    \label{fig:r_capture}
\end{figure}

Panel~(a) in Fig.~\ref{fig:r_capture} presents the cumulative mass distribution of captured planetesimals for different times as indicated in the legend. Although ablation occurs throughout the CPD, capture occurs primarily at distances of $\gtrsim 10R_\mathrm{Jup}$. Captured planetesimals may serve as seed satellitesimals onto which pebbles produced from the ablated material subsequently accrete. Because efficient ablation occurs primarily within $\lesssim10\,R_\mathrm{Jup}$, only a small fraction of the ablated material would be available for satellitesimal growth if the ablated solids drift only inward.

Planetesimal capture occurs primarily during the early stage of Jupiter's formation, at $t-t_0\lesssim 3\times10^{5}~\mathrm{yr}$. As proto-Jupiter grows, the capture region shifts to larger distances because the CPD expands with the planetary growth and gas drag enables incoming planetesimals to lose sufficient orbital energy for capture at larger radii.

Panel~(b) shows the radii of the captured planetesimals. During the orbital damping that leads to capture by the CPD, planetesimals lose mass through ablation. Nevertheless, most captured planetesimals retain radii close to their initial values.

\subsection{Dependence on the planetesimal size} 

As shown in Fig.~\ref{fig:orbs}, the strength of gas drag affects the orbital evolution of planetesimals in the protoplanetary disk. In addition, both gas drag and ablation within the CPD depend strongly on planetesimal size. In this section, we investigate how the delivery of solid material to the CPD varies with the initial planetesimal radius.

\begin{figure}
    \centering
    \includegraphics[width=1.0\linewidth]{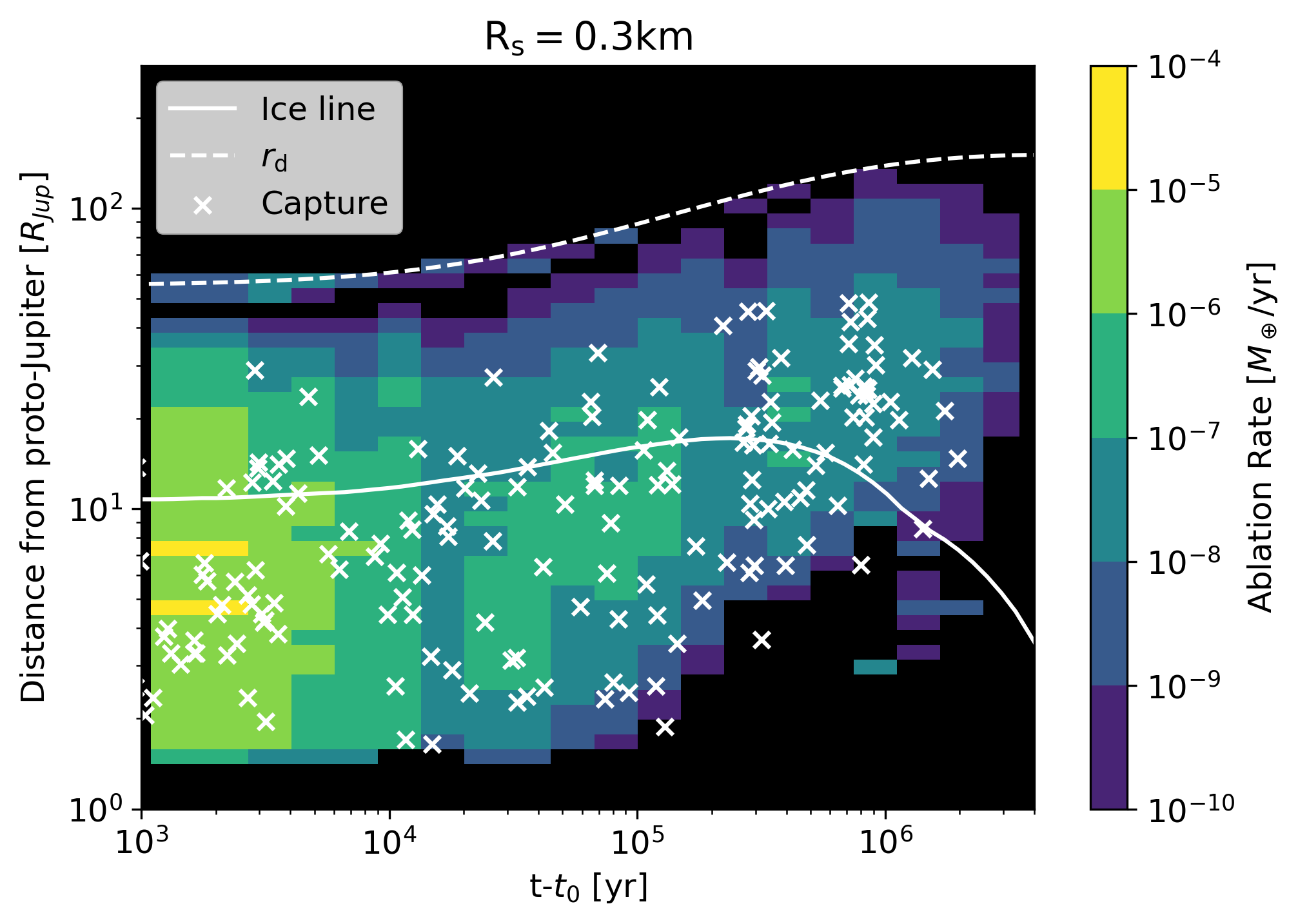}
    \caption{
    Same as Fig.~\ref{fig:t_r_ablation}, but for $R_\mathrm{s}=300$ m. 221 super-particles are captured in this case.
    }
    \label{fig:t_r_ablation_Rpl300m}
\end{figure}

Figure~\ref{fig:t_r_ablation_Rpl300m} provides an overview of material delivery to the CPD, analogous to Fig.~\ref{fig:t_r_ablation}, but for smaller planetesimals with $R_\mathrm{s}=300~\mathrm{m}$. The resulting ablation rate is higher than that obtained for planetesimals with $R_\mathrm{s}=100~\mathrm{km}$. The ablation timescale is given as
\begin{align}\label{eq:ablation_timescale}
    \tau_\mathrm{abl}&=\frac{M_\mathrm{pl}}{\left|\mathrm{d}M_\mathrm{pl}/\mathrm{d}t\right|}, \nonumber\\
    &= \frac{\rho_\mathrm{s} R_\mathrm{s}}{3 P_\mathrm{v} \left( T_\mathrm{s} \right)} \sqrt{\frac{2 \pi k_\mathrm{B} T_\mathrm{s}}{\mu_\mathrm{s}m_\mathrm{H}}}.
\end{align}
Therefore, smaller planetesimals more quickly evapolates in the CPD. 

The location of planetesimal ablation also shifts outward for smaller planetesimals. Equation~\ref{eq:ablation_timescale} shows that smaller planetesimals can be quickly ablated even in relatively low-temperature regions.
Consequently, a substantial amount of material is ablated outside the water ice line, particularly during the late stage of Jupiter's formation, at $t-t_0\gtrsim10^6~\mathrm{yr}$. This late delivery of icy material is favorable for the formation of ice-rich satellites such as Ganymede and Callisto.

We also find that smaller planetesimals are more readily captured by the CPD because their aerodynamic damping timescales are shorter, as shown in Eq.~\ref{eq:damping_timescale}. Since both the ablation and damping timescales decrease with planetesimal size, smaller planetesimals interact more efficiently with the disk gas, enhancing the delivery of material to the CPD through both ablation and direct capture. Contorary, when the planetesimals are large as $R_\mathrm{s}=100$ km, many planetesimals collide to proto-Jupiter before completely evapolating or being captured by the CPD. 

\begin{figure}
    \centering
    \includegraphics[width=0.9\linewidth]{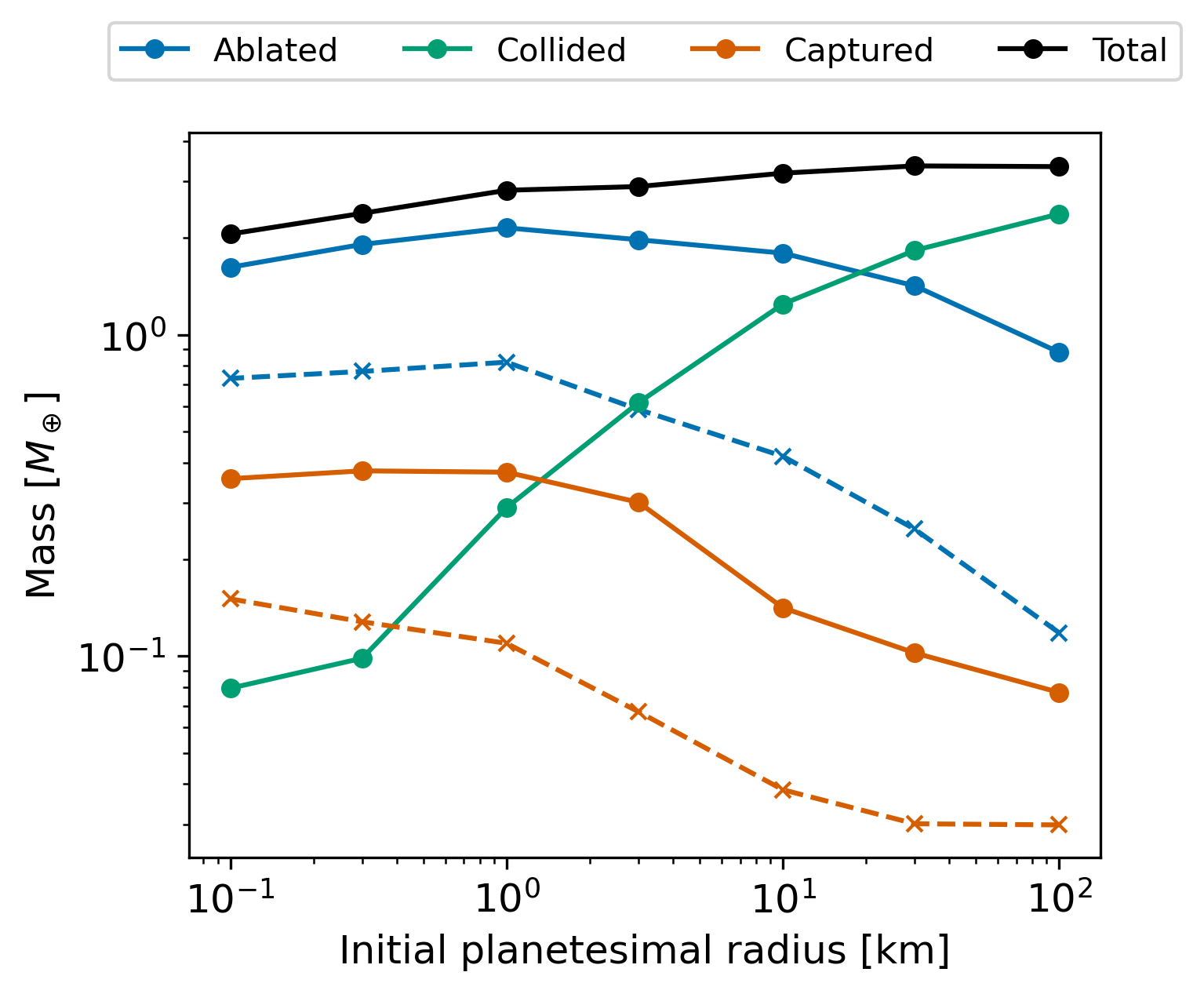}
    \caption{
    Mass of solid materials delivered to the Jovian system during Jupiter's formation. The blue, green, and orange circles indicate the masses of ablated planetesimals, planetesimals that collided with proto-Jupiter, and planetesimals captured by the CPD, respectively. The crosses show the corresponding contributions occurring outside the water ice line. The black circles indicate the total mass of planetesimals delivered to the Jovian system.
    }
    \label{fig:Rpl_mass}
\end{figure}

Figure~\ref{fig:Rpl_mass} shows the mass of solid materials delivered to Jovian system as a function of initial planetesimal radius. The blue, green, and orange plots show the total mass lost by ablation, collision, and capture, respectively. The black plots show the total mass lost in Jovian system. As shown in Fig.~\ref{fig:orbs}, larger planetesimals can stay in the feeding zone for longer time and have more chances to enter the CPD. Therefore, the total mass increases with planetesimal radius. 

The dominant delivery channel is either ablation or collision, depending on the initial planetesimal radius. By contrast, the mass directly captured by the CPD accounts for approximately $10\%$ of the total mass delivered to the Jovian system. The relative contributions of ablation and capture depend only weakly on planetesimal size because the ratio of the ablation timescale to the orbital-damping timescale is independent of the planetesimal radius.

Although more than $1~M_\oplus$ of solid material is delivered to the CPD, the amount available for building icy satellites is more limited. The blue and orange crosses in Fig.~\ref{fig:Rpl_mass} indicate the masses of material ablated and captured outside the water ice line, respectively. The total mass available for the formation of icy satellites is less than $1~M_\oplus$. This provides an important constraint on the formation of Jovian satellites, which we discuss in the following section.

\section{Hybrid accretion of satellite formation}\label{sec:satellite_formation}

From our numerical simulations above, we find that, for a planetesimal reservoir comparable in mass to that of the minimum-mass solar nebula, $f_\mathrm{PPD}=1$, a few $M_\oplus$ of solid material can be delivered to the CPD during Jupiter's formation through ablation and direct capture. The ablated material can supply a pebble flux, while directly captured planetesimals can form a planetesimal, or satellitesimal, ring from which the Jovian satellites may grow. In this section, we discuss how the satellites can accrete from these delivered solids and estimate the minimum mass of the planetesimal reservoir in the protoplanetary disk required for Jovian system formation.

\subsection{Onset of satellite formation}

The delivery of solid material to the CPD continues throughout Jupiter's formation. However, not all of the delivered material is necessarily available for building the Jovian satellites. Material delivered during the early stages of Jupiter's formation may subsequently be lost through inward migration toward proto-Jupiter. \citet{Canup2006} demonstrated that growing protosatellites can undergo rapid inward migration and eventually collide with proto-Jupiter. Because the accretion timescale is regulated by the local orbital period, which is much shorter in the inner CPD than the characteristic formation timescale of proto-Jupiter, this cycle of satellite growth and loss can proceed on a timescale much shorter than Jupiter's overall formation timescale, $<10^5~\mathrm{yr}$ \citep{Canup2006, Sasaki2010}.

A central cavity created by proto-Jupiter’s magnetic field may halt the inward migration of growing satellites and prevent their accretion onto the planet \citep{Sasaki2010, Ogihara2012, Shibaike2019, Madeira2021}. The formation of such a cavity is governed by the balance between the magnetic stress and the ram pressure of the accreting gas \citep[e.g.][]{Armitage2010}. Consequently, a magnetospheric cavity becomes more likely to open during the late stages of Jupiter’s growth, as the gas accretion rate declines \citep{Shibaike2019}.

Hereafter, we assume that the cavity opens at $t-t_0=10^5~\mathrm{yr}$, by which time proto-Jupiter has acquired a massive envelope, $>30~M_\oplus$, and may begin to generate a strong magnetic field.

\subsection{Required mass for building Gallilean satellites}

Pebble accretion captures only a fraction of the inward-drifting pebble flux. Therefore, forming a satellite of a given mass generally requires an integrated pebble flux substantially larger than the final satellite mass. Here, we estimate $M_\mathrm{req}$, the integrated pebble flux required to form Ganymede, the most massive icy satellite in the Jovian system, with a mass of $M_\mathrm{Gany}=0.0248~M_\oplus$ and an orbital radius of $r_\mathrm{Gany}=15.0~R_\mathrm{Jup}$.

\begin{figure}
    \centering
    \includegraphics[width=0.8\linewidth]{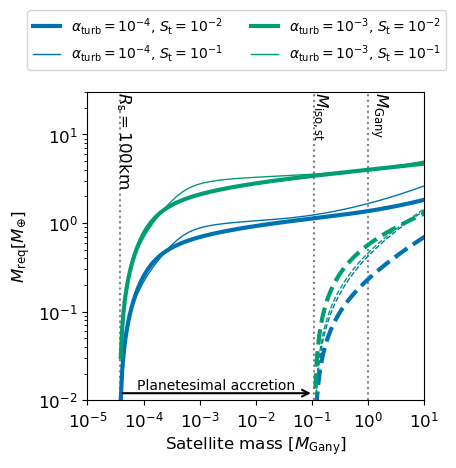}
    \caption{
    Integrated pebble flux required to form satellites, plotted as a function of the final satellite mass. The blue and green lines correspond to $\alpha_\mathrm{turb}=10^{-4}$ and $10^{-3}$, respectively. 
    \revise{
    The accretion viscosity parameter, $\alpha_\mathrm{CPD}$, is fixed in these calculations.
    }
    The thick and thin lines indicate cases with $\mathrm{St}=0.01$ and $0.1$, respectively. The solid lines are calculated assuming that pebble accretion begins from captured planetesimals with radii of $100~\mathrm{km}$, whereas the dashed lines assume initial protosatellite masses equal to the isolation mass attained through satellitesimal accretion. Here, we assume that the protosatellite orbits at $15~R_\mathrm{Jup}$.
    }
    \label{fig:mass_Mreq}
\end{figure}

We consider in situ formation for simplicity. For a pebble accretion efficiency $\epsilon$, $M_\mathrm{req}$ can be written as
\begin{align}\label{eq:M_req_satelite}
    M_\mathrm{req} = \int_{M_\mathrm{sat,0}}^{M_\mathrm{sat}} \frac{d m}{\epsilon}.
\end{align}
$M_\mathrm{sat,0}$ is the initial mass of seed satelite. We integrate Eq.~\ref{eq:M_req_satelite} using the pebble-accretion efficiency derived by \citet{Liu2018} and \citet{Ormel2018}. 
\revise{
This prescription accounts for the suppression of pebble
accretion onto low-mass seeds through the settling
modulation factor, $f_\mathrm{set}$, which decreases
rapidly when the seed mass becomes too small for
efficient settling accretion.
}

We consider two limiting cases. In the first case, pebble accretion begins directly onto individual captured planetesimals, without significant prior growth through mutual collisions. For $R_\mathrm{s}=100~\mathrm{km}$, the corresponding initial satellite-seed mass is $M_\mathrm{sat,0}=4\times10^{-5}M_\mathrm{Gany}$.

In the second case, satellite seeds undergo hybrid growth: they first grow through mutual collisions among captured satellitesimals and subsequently accrete pebbles. Our numerical results show that approximately $0.1~M_\oplus$ of planetesimals are captured by the CPD, forming a circumplanetary satellitesimal ring. As discussed in \ref{App:satelitesimal_ring}, the surface density of this ring can be approximated as
\begin{align}
    \Sigma_\mathrm{st}\sim2000~\mathrm{g/cm^2}\,f_\mathrm{PPD}
    \left(\frac{d}{10R_\mathrm{Jup}}\right)^{-1/2}.
\end{align}
Protosatellites can grow toward their local isolation mass,
\begin{align}\label{eq:M_iso_st}
    M_\mathrm{iso,st}
    =&\ 0.04M_\mathrm{Gany}
    \left(\frac{b}{10}\right)^{3/2}
    \left(\frac{M_\mathrm{p}}{M_\mathrm{Jup}}\right)^{-1/2}
    \nonumber\\
    &\times
    \left(\frac{\Sigma_\mathrm{st}}{2000~\mathrm{g/cm^2}}\right)^{3/2}
    \left(\frac{r}{10R_\mathrm{Jup}}\right)^3,
\end{align}
where $b$ is the radial width of the feeding zone in units of the mutual Hill radius during oligarchic growth. In this case, we integrate Eq.~\ref{eq:M_req_satelite} using $M_\mathrm{sat,0}=M_\mathrm{iso,st}$.

Figure~\ref{fig:mass_Mreq} shows the resulting $M_\mathrm{req}$ as a function of the satellite mass, $M_\mathrm{sat}$.
\revise{
Here, we calculate the pebble accretion efficiency, $\epsilon$, assuming
}
a sub-Keplerian velocity parameter of $\eta=0.01$, and a CPD aspect ratio of $H_\mathrm{CPD}/d=0.1$. We treat the Stokes number and a turbulent viscosity parameter, $\alpha_\mathrm{turb}$, as free parameters. 
\revise{
Note that we distinguish the turbulent viscosity parameter, $\alpha_\mathrm{turb}$, from the accretion viscosity parameter, $\alpha_\mathrm{CPD}$, and assume $\alpha_\mathrm{turb}\leq\alpha_\mathrm{CPD}$.
}
The Stokes number of drifting pebbles is expected to range from $10^{-2}$ to $10^{-1}$ \citep{Shibaike2019}.

We find that if satellites are formed via pebble accretion only, the integrated pebble flux required to form Ganymede is estimated to exceed $1~M_\oplus$. The ablated mass of planetesimlas outside the water ice line is less than that (see Fig.~\ref{fig:Rpl_mass}), implying that the planetesimal reservoir must have been more massive than that of the minimum-mass solar nebula.

In the hybrid growth scenario, however, satellitesimal accretion first produces massive protosatellite seeds, which subsequently grow through pebble accretion. The required integrated pebble flux is reduced by a factor of approximately $5$--$10$, as shown by the dashed lines in Fig.~\ref{fig:mass_Mreq}.

\subsection{Minimum mass of planetesimal resorvior}

Both the satellitesimal isolation mass and the pebble flux available in the CPD depend on the mass of the planetesimal reservoir in the protoplanetary disk, because both are ultimately set by the ablation and direct capture of planetesimals. By varying $f_\mathrm{PPD}$, we estimate the minimum planetesimal reservoir mass, or equivalently the minimum value of $f_\mathrm{PPD}$, required to form Ganymede.

\begin{figure}
    \centering
    \includegraphics[width=0.8\linewidth]{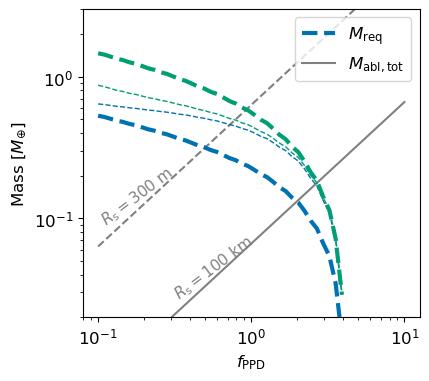}
    \caption{
        Integrated pebble flux required to \revise{form} Ganymede \revise{in our hybrid accretion model}, plotted as a function of the scaling factor for the planetesimal reservoir in the protoplanetary disk, $f_\mathrm{PPD}$. The colored lines show $M_\mathrm{req}$ estimated using Eq.~\ref{eq:M_req_satelite}. The colors and line thicknesses have the same meanings as in Fig.~\ref{fig:mass_Mreq}. The gray lines show the pebble mass available from planetesimal ablation in our $N$-body simulations. The solid and dashed gray lines correspond to initial planetesimal radii of $R_\mathrm{s}=100~\mathrm{km}$ and $300~\mathrm{m}$, respectively.
    }
    \label{fig:fppd_Mreq}
\end{figure}

In Fig.~\ref{fig:fppd_Mreq}, we plot the required integrated pebble flux, $M_\mathrm{req}$, in the hybrid-accretion scenario as a function of the planetesimal-disk scaling factor, $f_\mathrm{PPD}$. The required pebble mass decreases with increasing $f_\mathrm{PPD}$ and becomes zero for $f_\mathrm{PPD}>4$, because the satellitesimal isolation mass then exceeds the mass of Ganymede. In this regime, Ganymede can form through satellitesimal accretion alone.

We also plot the available integrated pebble flux, $M_\mathrm{abl,tot}$, with gray lines. Assuming that the ablated material forms pebbles that drift inward, $M_\mathrm{abl,tot}$ at $15~R_\mathrm{Jup}$ is given by
\begin{align}
    M_\mathrm{abl,tot} = \int F_\mathrm{abl} \left( r>15 R_\mathrm{Jup}, t-t_0 > 10^5 \mathrm{yr} \right) \mathrm{d} t \mathrm{d} r,
\end{align}
where $F_\mathrm{abl}$ is the planetesimal ablation rate obtained from our numerical simulations.

For planetesimals with radii of $100~\mathrm{km}$, $M_\mathrm{abl,tot}$ is less than $0.1f_\mathrm{PPD}~M_\oplus$ because relatively little ablation occurs outside the water ice line. In this case, forming Ganymede requires a planetesimal reservoir approximately two to three times more massive than that of the minimum-mass solar nebula, and its growth may be dominated by satellitesimal accretion. By contrast, for smaller planetesimals with radii of $300~\mathrm{m}$, ablation supplies pebbles more efficiently. Consequently, hybrid accretion can form Ganymede even from a planetesimal reservoir less massive than that of the minimum-mass solar nebula.

In the above estimate, we neglect dust transport associated with the accreting gas from the protoplanetary disk to the CPD. Although the dust-to-gas mass-flux ratio is expected to be as low as $\sim0.1\%$ for a nominal turbulent viscosity parameter of $\alpha_\mathrm{turb}=10^{-3}$ in the protoplanetary disk \citep{Maeda2024}, this inflow could still deliver approximately $0.3~M_\oplus$ of dust to the CPD. This additional source would increase the pebble mass available for satellite formation and further reduce the required mass of the planetesimal reservoir in the protoplanetary disk.

We also do not account for pebble filtering by multiple protosatellites or for their subsequent growth through giant impacts induced by convergent migration. The estimated satellitesimal isolation masses suggest that tens of protosatellites may form within the satellitesimal ring. The presence of multiple pebble-accreting bodies increases the system-wide pebble-accretion efficiency \citep{Guillot2014, Madeira2021}. \citet{Madeira2021} reported a system-wide pebble-accretion efficiency of approximately $9\%$ with their N-body simulations. Using the same model setup as \citet{Madeira2021}, Equation~\ref{eq:M_req_satelite} yields an accretion efficiency of approximately $2\%$ for a single protosatellite. This comparison suggests that pebble filtering by multiple protosatellites can enhance the overall accretion efficiency by a factor of a few. Consequently, the Galilean satellites may be able to form from a planetesimal reservoir with an even smaller value of $f_\mathrm{PPD}$ than estimated in our single-protosatellite calculation.

\section{Discussion}\label{sec:Discussion}

\subsection{Imprints on Solar System architectures}
\label{sec:discussion_solar_system}

By quantifying solid delivery to the CPD, we constrain the minimum mass of the planetesimal reservoir required to form the Galilean satellites, finding $f_\mathrm{PPD}\gtrsim0.5$. Although our primary focus is the delivery of solids to the CPD, the growing proto-Jupiter also accretes and scatters planetesimals, redistributing solid material into Jupiter’s deep interior and the main asteroid belt. These outcomes provide complementary constraints on Jupiter’s formation and, more broadly, the early evolution of the Solar System.

In this section, we compare our numerical results with other observational and theoretical constraints on Solar System formation.

\subsubsection{Depleted asteroid belt}

While approximately $20$--$30\%$ of the planetesimals near proto-Jupiter's orbit are delivered to the Jovian system (see Figure \ref{fig:Rpl_mass}), the remaining planetesimals stay in the protoplanetary disk. Most of these bodies pile up near the boundaries of Jupiter's feeding zone and may subsequently be ejected from the Solar System during the post-disk dynamical instability of the giant planets \citep{Dones2004}. A fraction of them, however, are scattered onto more interior orbits and become implanted into the main asteroid belt.

The asteroid belt is known to be strongly depleted. The estimated total mass of asteroids between $2.06$ and $3.27~\mathrm{au}$ is $(4.008\pm0.029)\times10^{-4}~M_\oplus$ \citep{Pitjeva2018}. The primordial asteroid belt may have been substantially more massive, with subsequent dynamical instability and collisional fragmentation depleting much of its original mass. Accretion and fragmentation models for the main asteroid belt suggest that its primordial mass was approximately $5$--$200$ times greater than its present-day value \citep{Bottke2005, Deienno2024}, corresponding to a maximum primordial mass of $\sim0.1~M_\oplus$.

We analyze the mass implanted into the main asteroid belt, defined here as $2.06~\mathrm{au}<r<3.27~\mathrm{au}$. We find that the implanted mass ranges $0.2$--$0.5 f_\mathrm{PPD} M_\oplus$, which is lower than the mass ablated in the CPD but comparable to, or higher than, the mass captured by the CPD. The present-day structure of the asteroid belt therefore places an upper limit on the initial planetesimal reservoir, suggesting $f_\mathrm{PPD}\lesssim0.2$--$0.5$. This upper limit only marginally overlaps with the minimum reservoir mass required to form the Galilean satellites obtained in our model.

If the primordial asteroid belt was substantially less massive than $0.1~M_\oplus$, as suggested by the accretion-fragmentation model of \citet{Deienno2024}, then $f_\mathrm{PPD}$ would need to be significantly smaller than 0.5. Such a low-mass planetesimal reservoir would be insufficient to deliver the amount of solid material required to form the Jovian satellite system.

\begin{figure}
    \centering
    \includegraphics[width=0.9\linewidth]{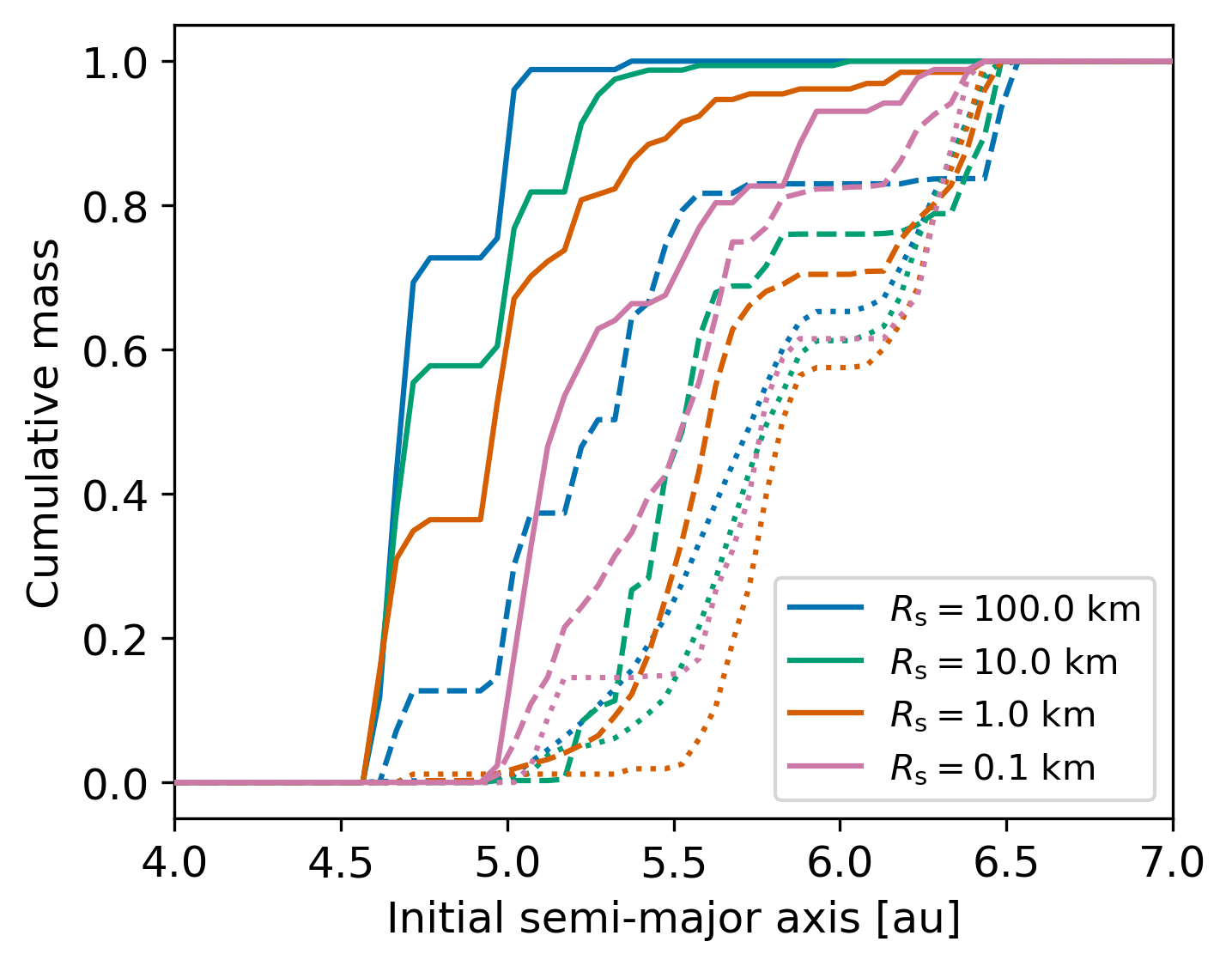}
    \caption{
    Normalized cumulative mass distributions as functions of the initial semimajor axis of the planetesimals. The solid, dashed, and dotted lines show the distributions of implanted, captured, and collided planetesimals, respectively. Different colors correspond to simulations with different initial planetesimal sizes. 
    }
    \label{fig:r0_asteroid}
\end{figure}

A possible solution is a different initial radial distribution of planetesimals. In our simulations, planetesimals are initially distributed over a broad region around proto-Jupiter. If Jupiter instead formed within a localized planetesimal ring and began runaway gas accretion near its inner edge, the efficiency of implantation into the asteroid belt could be substantially reduced. Figure~\ref{fig:r0_asteroid} shows that implanted planetesimals originate systematically closer to the Sun than those captured by the CPD or accreted by proto-Jupiter. Therefore, a radially concentrated planetesimal distribution, as suggested by recent planetesimal-formation models \citep[e.g.][]{Drazkowska2017}, could supply sufficient solid material to the CPD while avoiding excessive implantation into the asteroid belt.

In addition, planetesimals implanted from the region near Jupiter tend to retain high eccentricities. Such dynamically excited bodies may contribute more strongly to collisional fragmentation than to the accretional growth of main-belt objects, potentially relaxing the upper limit on $f_\mathrm{PPD}$. This possibility should be investigated in future studies.

\subsubsection{Jupiter's dilluted core}

Most of the heavy-element material delivered to the Jovian system ultimately accretes onto Jupiter, except for the fraction retained in the satellite system. Accreted planetesimals can modify the density and compositional structure of Jupiter's deep envelope, which can in turn be constrained by measurements of its gravitational moments from spacecraft observations.

Jupiter’s gravitational moments indicate the presence of a compositional gradient in its deep envelope, commonly referred to as a diluted core \citep[e.g.,][]{Wahl2017,Debras2019,Nettelmann2021,Miguel2022,Helled2022,Howard2023}. Interior-structure models suggest that the diluted core extends over more than $20\%$ of Jupiter’s mass \citep{Howard2023}. Its origin remains actively debated, but one promising scenario involves the hybrid accretion of pebbles and planetesimals \citep{Venturini2020}. Pebble accretion can build a core massive enough to trigger runaway gas accretion, but may not by itself produce the observed extended diluted core because pebble accretion ceases near the pebble-isolation mass, potentially truncating the compositional gradient \citep{Ormel2021}. By contrast, planetesimal accretion can continue to later stages and may produce a more extended diluted core \citep{Lozovsky2017,Stevenson2022}.

In our numerical simulations, more than half of the material delivered to the Jovian system is supplied before proto-Jupiter reaches $20\%$ of its present-day mass.
\revise{
This result is consistent with the findings of \citet{Shibata2022c}.
}
For $f_\mathrm{PPD}=1$, this corresponds to approximately $1$--$2~M_\oplus$ of planetesimals that could contribute to the formation of the diluted core. Because the resulting structure of the diluted core depends on the coupled histories of pebble, planetesimal, and gas accretion, it remains unclear what planetesimal-reservoir mass, or equivalently what value of $f_\mathrm{PPD}$, is required to reproduce the observed compositional gradient. Coupling planetesimal scattering and accretion calculations with models of Jupiter's interior and compositional evolution will therefore be important for placing an independent constraint on $f_\mathrm{PPD}$.

\subsection{Caveats on the Assumed Ablation Efficiency}

The ablation model adopted in this study follows previous studies \citep{DAngelo2015,Ronnet2020}. These works demonstrated that planetesimal ablation can make a significant contribution to the delivery of solid material to the CPD under idealized assumptions for the ablation process. Here, we assess how sensitive the ablation efficiency is to the adopted model assumptions.

\subsubsection{Heating efficiency}\label{sec:discussion_Ch}

\begin{figure}
    \centering
    \includegraphics[width=0.9\linewidth]{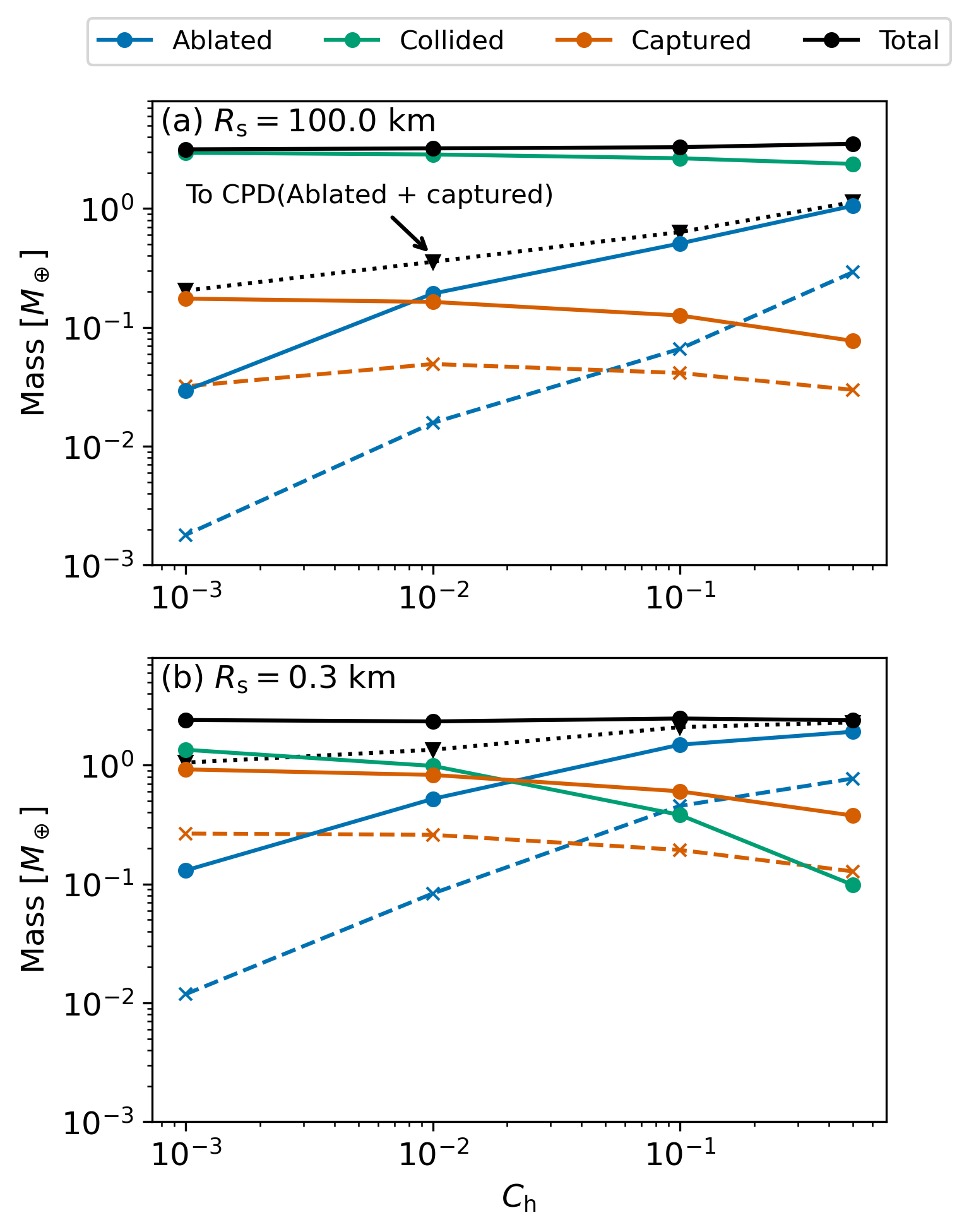}
    \caption{
    Same as Fig.~\ref{fig:Rpl_mass}, but shown as a function of the heat transfer coefficient, $C_\mathrm{h}$. The dotted line shows the mass delivered to the satellite system (ablated + captured). The top and bottom panels correspond to large planetesimals with $R_\mathrm{s}=100~\mathrm{km}$ and small planetesimals with $R_\mathrm{s}=300~\mathrm{m}$, respectively. The reference case, for which $C_\mathrm{h}=C_\mathrm{d}/4$, is plotted at $C_\mathrm{h}=0.5$, because $C_\mathrm{d}\simeq2$ for planetesimal-sized bodies.
    }
    \label{fig:Ch_Mlost}
\end{figure}

The final fate of planetesimals is strongly influenced by the efficiency of aerodynamic heating due to gas drag in the CPD, which is parameterized by the heat-transfer coefficient $C_\mathrm{h}$. We adopted an upper-limit value of $C_\mathrm{h}=C_\mathrm{d}/4\simeq0.5$, following previous studies \citep{DAngelo2015,Ronnet2020}. However, meteor observations on Earth suggest lower values, $C_\mathrm{h}\simeq0.05$--$0.2$ \citep[][see also \ref{App:heat_transfer_coefficient}]{Ceplecha2005,Brykina2021}. Numerical studies have likewise inferred smaller values, including $C_\mathrm{h}=10^{-3}$--$10^{-2}$ \citep{Svetsov1995}, $C_\mathrm{h}<0.045$ \citep{Johnston2018}, and $C_\mathrm{h}=0.02$--$0.05$ \citep{Dias2020}. Thus, the appropriate value of $C_\mathrm{h}$  remains highly uncertain. 

To assess the influence of aerodynamic heating, we vary $C_\mathrm{h}$. Figure~\ref{fig:Ch_Mlost} shows the masses of planetesimals delivered to the Jovian system as a function of $C_\mathrm{h}$. We find that the ablated mass decreases for smaller $C_\mathrm{h}$. Contorary, the masses of captured and collided planetesimals is larger for smaller $C_\mathrm{h}$. 

For values of $C_\mathrm{h}=0.01$--$0.1$, as inferred from observations and numerical simulations, material delivery through ablation becomes less efficient, and direct planetesimal capture may instead dominate the transport of solids to the CPD. In this regime, the growth of the Jovian satellites would rely more strongly on satellitesimal accretion, while the contribution from solids produced by planetesimal ablation would be reduced. Constraining $C_\mathrm{h}$ is therefore essential for accurately quantifying the delivery of building blocks to satellite systems.

\subsubsection{Viscous-heating disk}\label{Sec:Viscous_Heating_Disk}

In the simulations above, we neglected viscous heating when calculating the CPD temperature profile, assuming that wind-driven accretion reduces midplane heating compared with classical viscously evolving disks. However, some level of internal dissipation may remain, depending on the vertical disk structure, ionization state, and mechanism of angular-momentum transport \citep{Suzuki2016,Mori2019,Lesur2021}. This uncertainty is particularly relevant for CPDs, where the efficiency of MRI-driven transport depends sensitively on the ionization structure \citep{Fujii2014}. Here, we investigate how viscous heating affects the capture and ablation of planetesimals by considering disk models that include both irradiation and viscous heating.

The temperature structure of a viscously heated disk depends on the opacity. We parameterize the dust opacity using $\kappa_\mathrm{dust,0}$, with larger values of $\kappa_\mathrm{dust,0}$ producing higher disk temperatures. The details of the viscously heated disk model are described in \ref{app:viscously_heating_disk}.

\begin{figure}
    \centering
    \includegraphics[width=0.9\linewidth]{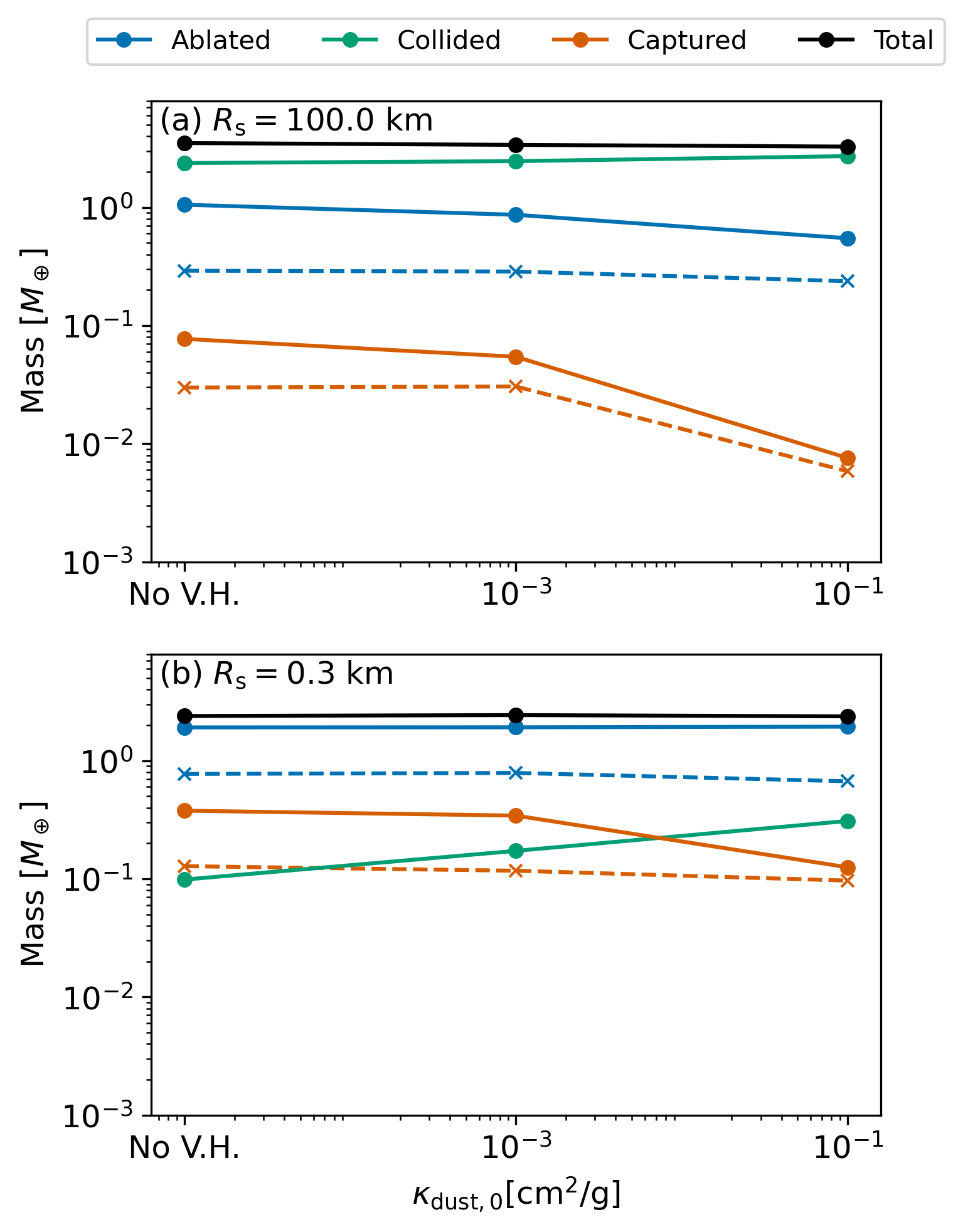}
    \caption{
    Same as Fig.~\ref{fig:Ch_Mlost}, but for simulations with viscous-heating disk shown as a function of the dust opacity factor, $\kappa_\mathrm{dust,0}$. We also show the results obtained without viscous heating, labeled ``No V.H.''.
    }
    \label{fig:Kappa_Mlost}
\end{figure}

Figure~\ref{fig:Kappa_Mlost} shows the results of our numerical simulations. In disks that include viscous heating, both the ablated mass and the mass captured by the CPD are smaller than in disks heated solely by irradiation. By contrast, the mass accreted by proto-Jupiter through collisions increases, while the total mass delivered to the Jovian system changes only weakly.

This result may appear counterintuitive because viscous heating increases the local disk temperature and might therefore be expected to enhance planetesimal ablation. However, the hotter midplane causes the CPD to expand vertically, thereby reducing the local gas density, $\rho_\mathrm{CPD}$. The planetesimal surface temperature, $T_\mathrm{s}$, is determined by heating due to gas drag and irradiation from the surrounding disk gas. Immediately after a planetesimal enters the CPD, aerodynamic heating dominates. In disks with viscous heating, the lower $\rho_\mathrm{CPD}$ weakens aerodynamic heating and results in a lower $T_\mathrm{s}$. Consequently, the ablation timescale, $\tau_\mathrm{abl}$, becomes longer.

\revise{
In addition to the viscous heating considered in our model, the kinetic energy carried by gas infalling from the PPD onto the CPD provides an additional heating source. The infalling gas generates an accretion shock near the CPD surface, where its kinetic energy is converted into thermal energy, followed by radiative cooling \citep{Aoyama2018}. Using radiation hydrodynamic simulations, \citet{Chen2026} showed that the infalling gas is heated to approximately 500 K at the shock front and subsequently cools to approximately 200 K in the post-shock region. In their simulations, the shock forms in the optically thin disk atmosphere, allowing efficient radiative cooling. Although radiation emitted from the shock-heated gas may contribute to heating the CPD midplane, \citet{Chen2026} showed that the midplane temperature profile obtained from their radiation hydrodynamic simulations can be well reproduced by a one-dimensional disk model accounting for viscous heating and radiative transport, without explicitly including shock heating. This agreement suggests that the contribution of infall-induced shock heating to the midplane temperature may be relatively small compared to viscous heating, although its effect has not been independently quantified.
}


Taken together with the uncertainty in the heat-transfer coefficient discussed in Sec.\ref{sec:discussion_Ch}, these results suggest that material delivery to the CPD may be less efficient under realistic ablation and CPD conditions than in our fiducial model. 
\revise{
Nevertheless, captured planetesimals may still play an important role for satellite formation. Even if the total mass of captured planetesimals were an order of magnitude smaller than in our fiducial model, mutual collisions among captured planetesimals could still produce satellitesimal seeds more than ten times more massive than the initially captured planetesimals. These larger seeds could enable efficient pebble accretion if additional solid material is supplied to the CPD through other delivery channels, such as dust entrained in the accreting gas.
}

\section{Summary and Conclusions}
\label{sec:summary}

\revise{
We investigated the delivery of solid material to the circumplanetary disk (CPD) during the growth of proto-Jupiter. Using $N$-body simulations combined with models of Jupiter's growth and CPD evolution, we followed the ablation and capture of planetesimals entering the Jovian system. Our main contribution from former studies \citep[e.g.,][]{, DAngelo2015, Suetsugu2016, Suetsugu2017, Ronnet2020} is to directly connect Jupiter's formation and the evolution of the surrounding planetesimal disk to the delivery of building blocks for the Galilean satellites. Our main conclusions are as follows. 

Although proto-Jupiter rapidly opens a gap in the planetesimal disk, its expanding and migrating feeding zone continues to supply planetesimals for more than $10^6~\mathrm{yr}$. This is significantly longer than the $\sim10^4~\mathrm{yr}$ evolution followed by \citet{Ronnet2020}, who assumed a fixed Jupiter mass and CPD structure. This prolonged supply may have important implications for the accretion and thermal histories of the Galilean satellites. 

As proto-Jupiter grows, the radial locations of planetesimal ablation and capture shift outward, while the water ice line moves inward as the CPD evolves. 
The relative evolution of the ablation/capture regions and the water ice line may affect the delivery of rocky and icy materials to different regions of the CPD, potentially influencing the compositions of the Galilean satellites.

In our reference model with $R_\mathrm{s}=100~\mathrm{km}$, planetesimal ablation supplies $0.88~M_\oplus$ of material to the CPD, whereas approximately $0.1~M_\oplus$ of planetesimals are directly captured. 
By varying the initial planetesimal size ($R_\mathrm{s}$) and the mass of the primordial planetesimal disk ($f_\mathrm{PPD}$), we quantitatively evaluate the conditions under which sufficient solid material can be delivered to the CPD. This connects the supply of satellite building blocks to the primordial structure of the planetesimal disk in the PPD.

Based on the structure of the satellitesimal ring formed by direct planetesimal capture, we find that mutual collisions and accretion among satellitesimals can produce large protosatellite seeds several orders of magnitude more massive than the initially captured planetesimals. Starting pebble accretion from these seeds reduces the pebble mass required to form Ganymede by a factor of approximately $5$--$10$. Hybrid accretion involving both satellitesimals and pebbles may therefore represent an important formation pathway for the Galilean satellites.

However, the amount of material delivered to the CPD depends sensitively on the assumptions regarding planetesimal ablation and the thermal structure of the disk. By varying the heat-transfer coefficient ($C_\mathrm{h}$) and the CPD heating model, we find that less efficient aerodynamic heating of planetesimals or stronger viscous heating of the CPD can suppress planetesimal ablation and increase the fraction of planetesimals that collide with proto-Jupiter. These results highlight the importance of uncertain physical processes in determining the efficiency of solid material delivery. In particular, the contribution of ablated planetesimals to the solid supply may be less efficient than predicted under the idealized conditions adopted in previous studies.

Overall, our study demonstrates that Jupiter's growth and the global orbital evolution of planetesimals play important roles in regulating the amount, timing, and location of solid material delivered to the Jovian satellite system. As suggested by previous studies, planetesimal ablation may supply sufficient material to form the Galilean satellites through pebble accretion, but this requires favorable conditions under which ablation is highly efficient. By contrast, direct planetesimal capture can provide sufficient mass to form a satellitesimal ring that hosts large satellite seeds. Combined with additional sources of solids, such as planetesimals scattered by other giant planets and dust delivered with the accreting gas, hybrid accretion involving satellitesimals and pebbles may have formed the Galilean satellites. This scenario may operate even if the planetesimal reservoir in the protoplanetary disk was less massive than that of the minimum-mass solar nebula.
}

\section*{CRediT authorship contribution statement}

Sho Shibata: Conceptualization, Methodology, Software,
Formal analysis, Investigation, Visualization,
Writing -- original draft.

Andr\'e Izidoro: Conceptualization, Supervision,
Writing -- review \& editing.

\section*{Acknowledgements}
We thank the anonymous referees for their careful reading of the manuscript and constructive comments and suggestions, which helped improve the clarity and quality of this work.
The numerical computations were carried out on PC cluster at the Center for Computational Astrophysics, National Astronomical Observatory of Japan.

\appendix

\section{Jupiter's growth model}\label{app:Jupiter_model}

We adopt Jupiter's formation model developed by \citet{Shibata2022c}. The gas accretion timescale, $\tau_\mathrm{acc}$, is given by \citep{Tanigawa2002}.
\begin{align}
    \tau_\mathrm{acc} = \frac{M_\mathrm{p}}{D \Sigma_\mathrm{PPD}}, \label{eq:GasAccretionTimescale}
\end{align}
with
\begin{align}
    D = 0.29 \left( \frac{M_{\rm p}}{M_{*}} \right)^{4/3} \left( \frac{H_{\rm PPD}}{r_{\rm p}} \right)^{-2} {r_{\rm p}}^2 \Omega_{\rm K, p},
\end{align}
where $r_\mathrm{p}$ is the radial distance between the central star and proto-Jupiter, $M_\mathrm{*}$ is the mass of central star set to the solar mass, $\Sigma_\mathrm{PPD}$ is the gas surface density of protoplanetary disk, $H_\mathrm{PPD}$ is the scale height of protoplanetary disk, and $\Omega_\mathrm{K, p}$ is the orbital angular velocity of proto-Jupiter. The planetary migration timescale is given by \citep[e.g.][]{Tanaka2002}
\begin{align}
    \tau_\mathrm{mig} =  \frac{1}{2 C_\mathrm{m}} \frac{M_{*}}{M_{\rm p}}
                                \frac{M_{*}}{{r_{\rm p}}^2\Sigma_\mathrm{PPD}} 
                                \left(\frac{H_{\rm PPD}}{r_{\rm p}} \right)^{2} \Omega^{-1}_{\rm K, p}, \label{eq:MigrationTimescale}
\end{align}
where $C_\mathrm{m}$ is a factor that depends on Lindblad and corotation torques and set to 1.6 in this study. 

Both the gas-accretion and migration timescales depend on the structure of the protoplanetary disk. The disk midplane temperature is set by irradiation from the central star and is given by
\begin{align}
    T_\mathrm{PPD} = 150 \mathrm{K} \left( \frac{r}{1 \mathrm{au}} \right)^{-3/7},
\end{align}
where $r$ is the distance from the central star. Under the viscous evolution, the surface density of disk gas, $\Sigma_\mathrm{SS}$, follows the self-similar-solution, given as
\begin{align}
    \Sigma_\mathrm{SS} =& \frac{(2-\gamma) M_{\rm disk,0}}{2 \pi {R_{\rm disk}}^{2}} \left( \frac{r}{R_{\rm disk}} \right)^{-\gamma} T^{-\frac{\frac{5}{2}-\gamma}{2-\gamma}} \exp \left\{ -\frac{1}{T} \left(\frac{r}{R_{\rm disk}}\right)^{2-\gamma} \right\}, \label{eq:self-similar-solution}
\end{align}
with
\begin{align}
    \gamma &= \frac{3}{2} + \frac{{\rm d} \ln T_\mathrm{PPD}}{{\rm d} \ln r} \\
    T &= 1 + \frac{t}{\tau_{\rm vis}}, \label{eq:normalized_time} \\
    \tau_{\rm vis} &= \frac{{R_{\rm disk}}^2}{\nu_{\rm d}}, \label{eq:viscous_characteristic_timescale}
\end{align}
where $M_{\rm disk,0}$ is the disk total mass at $t=0$, $R_{\rm disk}$ is a radial scaling length of the disk, $\tau_{\rm vis}$ is the characteristic viscous timescale,  and $\nu_{\rm d}$ is a disk gas viscosity at $r=R_{\rm disk}$. We use the $\alpha$-viscosity model \citep{Shakura+1973} and the disk gas viscosity is written as $\nu=\alpha_{\rm PPD} H^2_{\rm PPD} \Omega_\mathrm{K}$, where $\alpha_{\rm PPD}$ is the viscosity parameter, and $\Omega_\mathrm{K}$ is the orbital angular velocity.

The surface density profile of gaseous disk is altered by the gap opening around the planet, the gas accretion onto the planet, and the disk depletion. 
We include these effects and calculate the surface density profile of protoplanetary disk, $\Sigma_{\rm PPD}$, by: 
\begin{align}
    \Sigma_{\rm PPD} = f_{\rm gap} f_{\rm acc} f_{\rm dep} \Sigma_{\rm SS}, \label{eq:Sigma_gas}
\end{align}
where $f_{\rm gap}$ is the gap opening factor, $f_{\rm acc}$ is the gas accretion factor, and $f_{\rm dep}$ is the disk depletion factor.
For the gap opening factor, we adopt the empirically obtained model by \citet{Kanagawa2017}.
The gap structure changes with the radial distance from the planet $\Delta r=|r-r_{\rm p}|/r_{\rm p}$ and $f_{\rm gap}$ is written as a function of $\Delta r$ by: 
\begin{align}
    f_{\rm gap} =
        \begin{cases}
            \displaystyle{            
            \frac{1}{1+0.04K} 
            } &{\rm for}~ \Delta r < \Delta R_1, \\
            \displaystyle{            
            4.0 {K^{\prime} }^{-1/4} \Delta r -0.32
            } &{\rm for}~\Delta R_1 < \Delta r < \Delta R_2 , \\
            \displaystyle{            
            1
            } &{\rm for}~ \Delta R_2 < \Delta r,
        \end{cases} \label{eq:gap_structure}.
\end{align}
with
\begin{align}
    K           &= \left( \frac{M_{\rm p}}{M_{*}} \right)^2 \left( \frac{H_{\rm PPD}}{r_{\rm p}} \right)^{-5} {\alpha_{\rm PPD}}^{-1}, \\
    K^{\prime}  &= \left( \frac{M_{\rm p}}{M_{*}} \right)^2 \left( \frac{H_{\rm PPD}}{r_{\rm p}} \right)^{-3} {\alpha_{\rm PPD}}^{-1}, \\
    \Delta R_1  &= \left\{ \frac{1}{4(1+0.04K)} +0.08 \right\} {K^{\prime}}^{1/4}, \\
    \Delta R_2  &= 0.33 {K^{\prime}}^{1/4}.
\end{align}
In the disk region inner to the planet, the disk's surface density is reduced by the gas accretion onto the planet.
When the gas accretion timescale is given by eq.~(\ref{eq:GasAccretionTimescale}) and the gap structure is given by eq.~(\ref{eq:gap_structure}), $f_{\rm acc}$ is written as \citep{Tanaka2020}: 
\begin{align}
    f_{\rm acc} = 
        \begin{cases}
            1 &{\rm for}~r > r_{\rm p}, \\
            \displaystyle{            
            \left\{ 1 + \frac{D}{3\pi \nu (1+0.04K)} \right\}^{-1}
            } &{\rm for}~r \leq r_{\rm p}.
        \end{cases} \label{eq:sweet_spot}
\end{align}
To account for disk depletion processes, such as photo-evaporation or disk wind, we set the disk depletion factor $f_{\rm dep}$ to: 
\begin{align}
    f_{\rm dep} = \exp \left( - \frac{t}{\tau_{\rm dep}} \right).
\end{align}
The surface density of disk gas at the bottom of the gap, which is used for the gas accretion rate and migration rate, is obtained by  $\Sigma_{\rm PPD}=\Sigma_{\rm PPD} (r=r_{\rm p})$ using eq.~(\ref{eq:Sigma_gas}).

\begin{figure}
    \centering
    \includegraphics[width=0.9\linewidth]{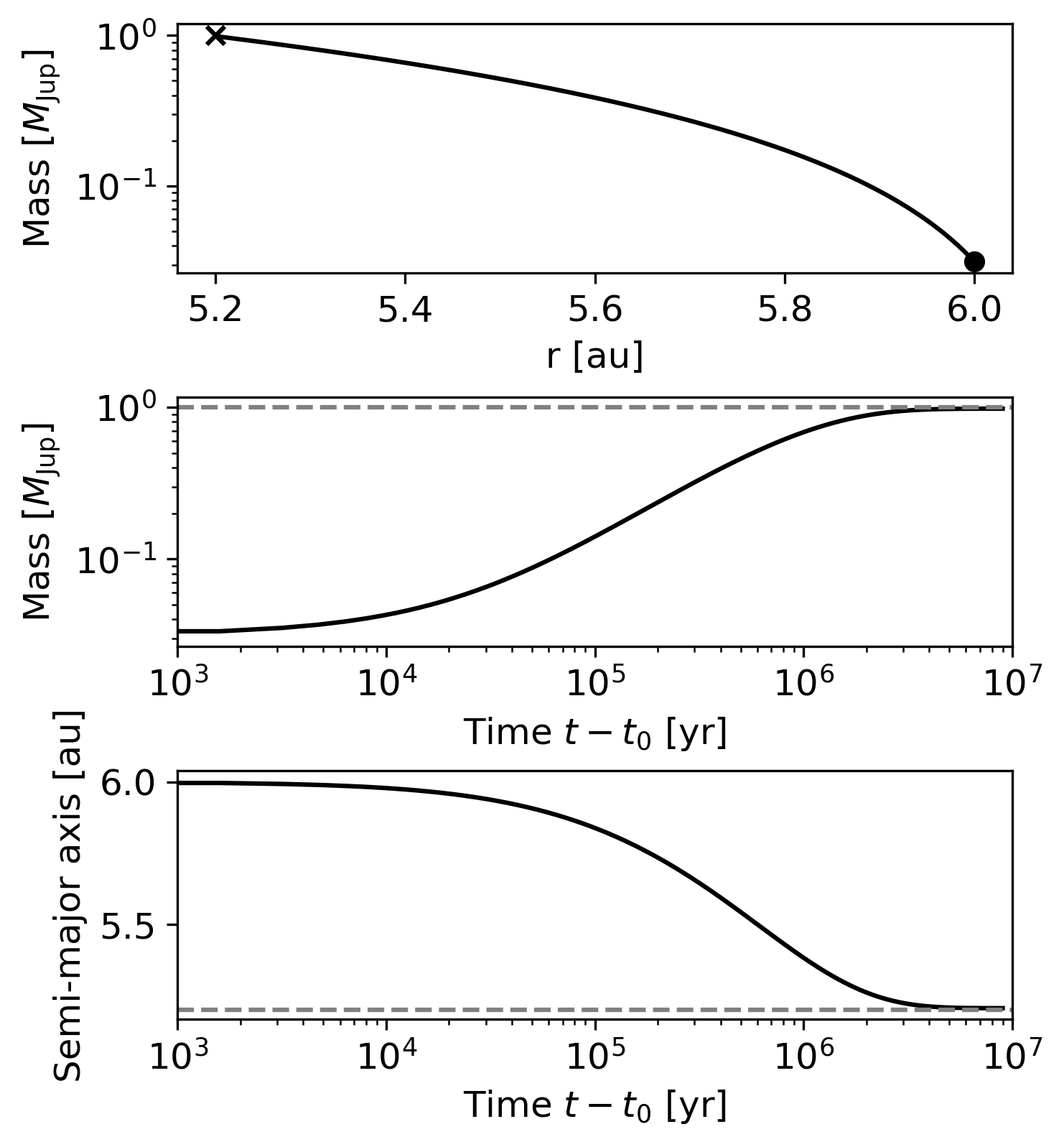}
    \caption{
    Growth track of proto-Jupiter obtained from our formation model. From top to bottom, the panels show the track in the $r_\mathrm{p}$--$M_\mathrm{p}$ plane and the time evolution of the planetary mass and semimajor axis, respectively.
    }
    \label{fig:Jupiter}
\end{figure}

The parameters governing Jupiter's formation are chosen such that the planet reaches approximately its present-day orbit and mass, and are summarized in Table~\ref{tb:model_Jupiter}. Using the equations above, we calculate the growth and migration of proto-Jupiter. Figure~\ref{fig:Jupiter} shows the resulting growth track and the timescale evolution.

\begin{table}
	\centering
	\begin{tabular}{ llllll } 
		\hline
		\hline
		$M_{*}$ 		& Mass of central star	                        & $1.0 M_\odot$ &	\\
		$M_{\rm disk,0}$    & Initial mass of protoplanetary disk           & $0.037 M_\odot$ &	\\
		$R_{\rm disk}$      & Typical size of protoplanetary disk           & $108 \mathrm{au}$ &	\\
		$\tau_{\rm dep}$ 	& Disk depletion timescale                      & $1\times10^6 \mathrm{yr}$ & \\
		$\alpha_\mathrm{PPD}$ 		& $\alpha$-viscosity parameter in PPD   & $10^{-3}$ &	\\
		$t_0$ 	& Time to start runaway gas accretion                      & $1\times10^6 \mathrm{yr}$ & \\
		$R_{\rm p}$ 	    & Physical radius of planet                   & $1.6 R_\mathrm{Jup}$ & \\
		\hline
	\end{tabular}
	\caption{
	Parameters used in Jupiter's formation model.
    }
    \label{tb:model_Jupiter}
\end{table}

\section{Aerodynamic drag coefficient}\label{app:Cd}

The drag coefficient, $C_{\rm d}$, depends on the flow regime around the planetesimal and is therefore determined by the Reynolds and Mach numbers, which depend on the planetesimal radius, relative velocity, and local gas properties. An approximated formula for $C_{\rm d}$ is given by: \citep[e.g.][]{Tanigawa2014} \begin{align}\label{eq:Drag_Coefficient_Tanigawa}
	C_{\rm d} \simeq \left[ \left( \frac{24}{\mathcal{R}} + \frac{40}{10+\mathcal{R}} \right)^{-1} + \frac{3 \mathcal{M}}{8} \right]^{-1} + \frac{ (2-\omega) \mathcal{M}}{1+\mathcal{M}} + \omega,
\end{align}
where $\mathcal{R}$ is the Reynolds number, $\mathcal{M}$ is the Mach number, and $\omega$ is a correction factor.
These parameters are given by: 
\begin{align}
    \mathcal{R} &= \frac{2 R_{\rm pl} u}{\nu_{\rm l}},\\
    \mathcal{M} &= \frac{u}{c_{\rm s}},\\
    \omega &= 
        \begin{cases}
                \displaystyle{ 0.4
                } &{\rm for}~\mathcal{R} < 2 \times 10^5, \\
                \displaystyle{ 0.2           
                } &{\rm for}~\mathcal{R} > 2 \times 10^5,
        \end{cases}
\end{align}
where $\nu_{\rm l}$ is the kinetic viscosity, and $c_\mathrm{s}$ is the sound speed of disk gas. For the ideal gas, $\nu_{\rm l}$ is obtained by: 
\begin{align}
    \nu_{\rm l} &= \frac{1}{3} \sqrt{\frac{8}{\pi}} c_{\rm s} l_{\rm p},
\end{align}
where $l_{\rm p}$ is the mean free path of a molecule in the disk gas.
Here, we use $l_{\rm p}=m_{\rm mol}/(\sigma_{\rm mol}\rho_{\rm gas})$, where $m_{\rm mol}$ and $\sigma_{\rm mol}$ are the molecular mass and collisional cross section, respectively.

\section{Vapor pressure model}\label{app:vaper_pressure}

\begin{table}
    \centering
    \begin{minipage}[t]{0.35\columnwidth}
        \centering
        \begin{tabular}{|c|c|}
            \hline
            Parameter & Value \\
            \hline
            $T_\mathrm{m}$ & $273.16~\mathrm{K}$ \\
            $a_0$ & $-2445.5646$ \\
            $a_1$ & $8.2312$ \\
            $a_2$ & $-0.01677006$ \\
            $a_3$ & $0.0000120514$ \\
            $a_4$ & $-6.757169$ \\
            \hline
        \end{tabular}
    \end{minipage}
    \hfill
    \begin{minipage}[t]{0.50\columnwidth}
        \centering
        \begin{tabular}{|c|c|}
            \hline
            Parameter & Value \\
            \hline
            $T_\mathrm{c}$ & $647.096~\mathrm{K}$ \\
            $P_\mathrm{c}$ & $22.064~\mathrm{MPa}$ \\
            $b_1$ & $-7.85951783$ \\
            $b_2$ & $1.84408259$ \\
            $b_3$ & $-11.7866497$ \\
            $b_4$ & $22.6807411$ \\
            $b_5$ & $-15.9618719$ \\
            $b_6$ & $1.80122502$ \\
            \hline
        \end{tabular}
    \end{minipage}
    \caption{Coefficients for the vapor pressure of water.}
    \label{tab:Pv_coeff}
\end{table}

The vapor pressure is obtained from the fitting functions available in literature. For icy bodies below the melting point ($T_\mathrm{s}<T_\mathrm{m}=273.16$ K), we use the formula of \citet{Washburn1924}, given as
\begin{align}
    \log \frac{P_\mathrm{v}}{\mathrm{mmHg}} = \frac{a_0}{T_\mathrm{s}} +a_1 \log_\mathrm{10} T_\mathrm{s} +a_2 {T_\mathrm{s}} +a_3 {T_\mathrm{s}}^2 +a_4,\\ 
\end{align}
where the pressure and temperature are in Ba and K. Above the melting point, we use the formula by \citet{Wagner2002}
\begin{align}
    \log\left(\frac{P_{\mathrm{v}}}{P_{\mathrm{c}}}\right)
    &=
    \frac{T_\mathrm{c}}{T_\mathrm{s}}
    \left(
    b_1\vartheta
    +b_2\vartheta^{1.5}
    +b_3\vartheta^3
    +b_4\vartheta^{3.5}
    +b_5\vartheta^4
    +b_6\vartheta^{4.5}
    \right), \\
    \vartheta
    &=1-\frac{T_\mathrm{s}}{T_{\mathrm{c}}},
\end{align}
where $T_\mathrm{c}$ and $P_\mathrm{c}$ are critical temperature and pressure, respectively. Coefficients and parameters related to the vapor pressure model are summarised in Tab.~\ref{tab:Pv_coeff}.

\section{Structure of captured satelitesimal ring}\label{App:satelitesimal_ring}

\begin{figure}
\centering
\includegraphics[width=0.9\linewidth]{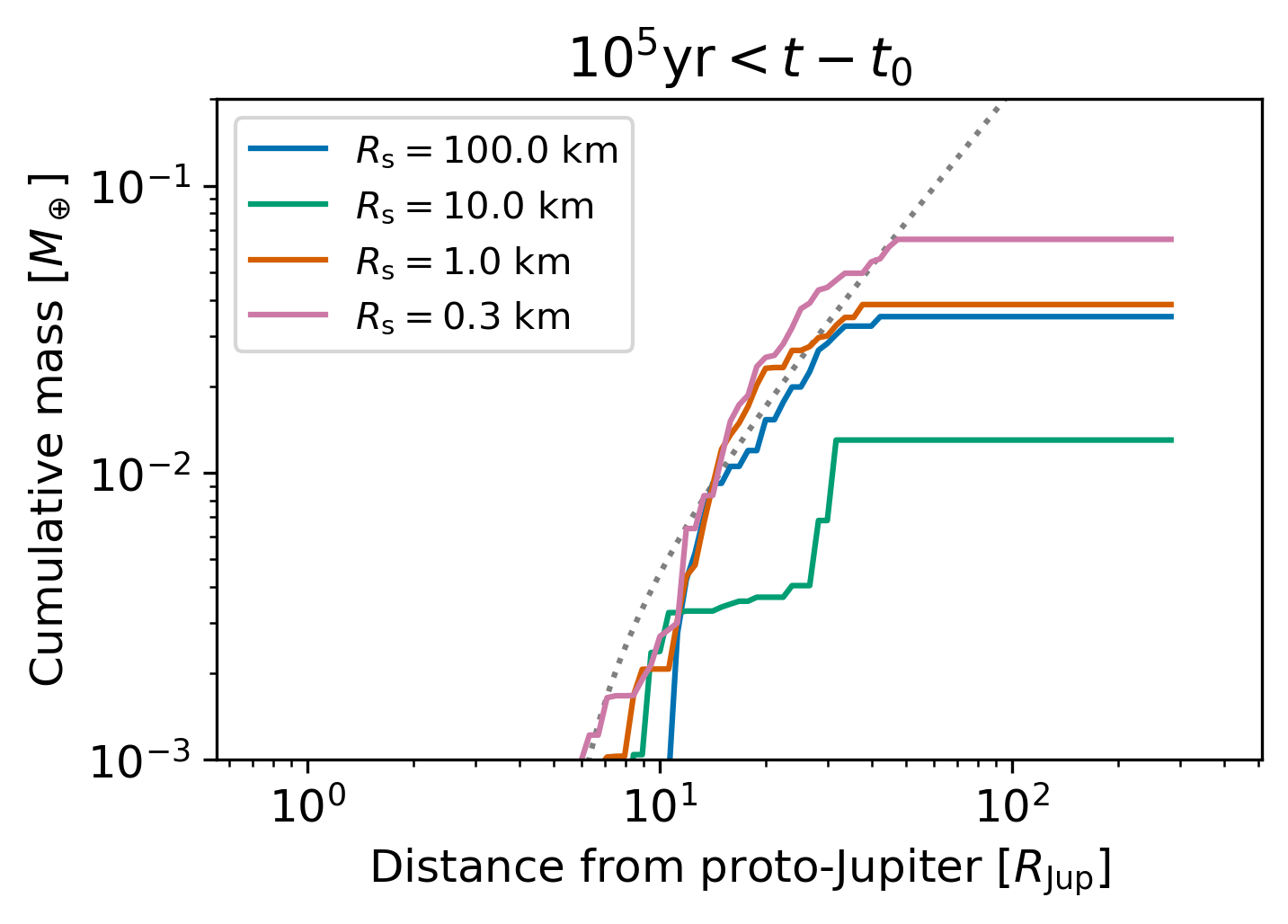}
\caption{
    Cumulative mass of captured planetesimals as a function of distance from proto-Jupiter. Different colors correspond to different initial planetesimal sizes. The gray dashed line shows the cumulative mass distribution of a satellitesimal ring with the surface density profile $\Sigma_\mathrm{st}=2000~\mathrm{g/cm^{2}} \left(d/10R_\mathrm{Jup}\right)^{-1/2}$.
}
\label{fig:r_capture_psRpl}
\end{figure}

Figure~\ref{fig:r_capture_psRpl} shows the cumulative radial mass distribution of captured planetesimals obtained in our numerical simulations. Here, we include only planetesimals captured after $t-t_0>10^5~\mathrm{yr}$. The captured planetesimals form a planetesimal, or satellitesimal, ring extending from approximately $5$ to $40~R_\mathrm{Jup}$. 
\revise{
We find that the resulting cumulative mass distribution is well reproduced by assuming a surface density profile of
$\Sigma_\mathrm{st} = 2000~\mathrm{g\,cm^{-2}}
\left(d/10R_\mathrm{Jup}\right)^{-1/2}$,
except for the case with $R_\mathrm{s}=10$ km.
}

\section{Estimated heat transfer coefficient}\label{App:heat_transfer_coefficient}

Based on precise observations of 15 meteorite-producing fireballs, \citet{Ceplecha2005} estimated typical intrinsic ablation coefficients of $\sigma_\mathrm{abl}=0.004$--$0.008~\mathrm{s^2~km^{-2}}$. The intrinsic ablation coefficient is related to the heat-transfer coefficient by \citep{Popova2006}
\begin{align}
    \sigma_\mathrm{abl} = \frac{C_\mathrm{h}}{2L_\mathrm{s}C_\mathrm{d}}.
\end{align}
Adopting $L_\mathrm{s}=6\times10^6~\mathrm{J~kg^{-1}}$ for an ordinary-chondrite-like composition \citep{Brykina2021} and $C_\mathrm{d}\simeq1$, this range corresponds to $C_\mathrm{h}\simeq0.05$--$0.10$.

\section{Viscously heating disk}\label{app:viscously_heating_disk}

The mid-plane temperature in the optically thick disk can be written as \citep{Nakamoto1994}
\begin{align}\label{eq:T_CPD_vis}
    2 \sigma_\mathrm{SB} T_\mathrm{vis}^4 = \frac{3}{8} \tau_\mathrm{R} \cdot \frac{9}{4} \Sigma_\mathrm{CPD} \nu \Omega_\mathrm{K}^2,
\end{align}
where $\tau_\mathrm{R}$ is the Rosseland mean optical depth. $\tau_\mathrm{R}$ can be written as $\kappa_\mathrm{disk} \Sigma_\mathrm{CPD}$ with the disk opacity $\kappa_\mathrm{disk}$. 

We model the disk opacity as a summation of the gas opacity, $\kappa_\mathrm{gas}$, and dust opacity, $\kappa_\mathrm{dust}$. We adopt the H$_2$ molecule's gas opacity from \citet{Bell1994} given by
\begin{align}\label{eq:kappa_disk_gas}
    \kappa_\mathrm{gas} = \kappa_\mathrm{gas,0} {\rho_\mathrm{gas}}^a {T_\mathrm{CPD}}^b,
\end{align}
where $\kappa_\mathrm{gas,0}=10^{-8} \mathrm{cm^2/g}$, and $a=2/3$, $b=3$. The dust opacity is a free parameter multiplied by the reduction factor due to the dust evapolation, which linearly interpolate $\kappa_\mathrm{dust,0}$ and 0 from 1,600 K to 2,000 K. 

\begin{table}
	\centering
	\begin{tabular}{ llllll } 
		\hline
		\hline
		$\kappa_\mathrm{gas,0}$ & gas opacity parameter & $10^{-8}$ cm$^2$/g\\
		$a$ & gas opacity parameter & 2/3 \\
		$b$ & gas opacity parameter & 3 \\
		$\kappa_\mathrm{dust,0}$ & dust opacity parameter & $0, 10^{-3}, 10^{-1}$ cm$^2$/g \\
		\hline
	\end{tabular}
	\caption{
    Parameters used in the viscous-heating CPD model.
    }
    \label{tb:model_viscous_heating_disk}
\end{table}

\begin{figure}
    \centering
    \includegraphics[width=0.9\linewidth]{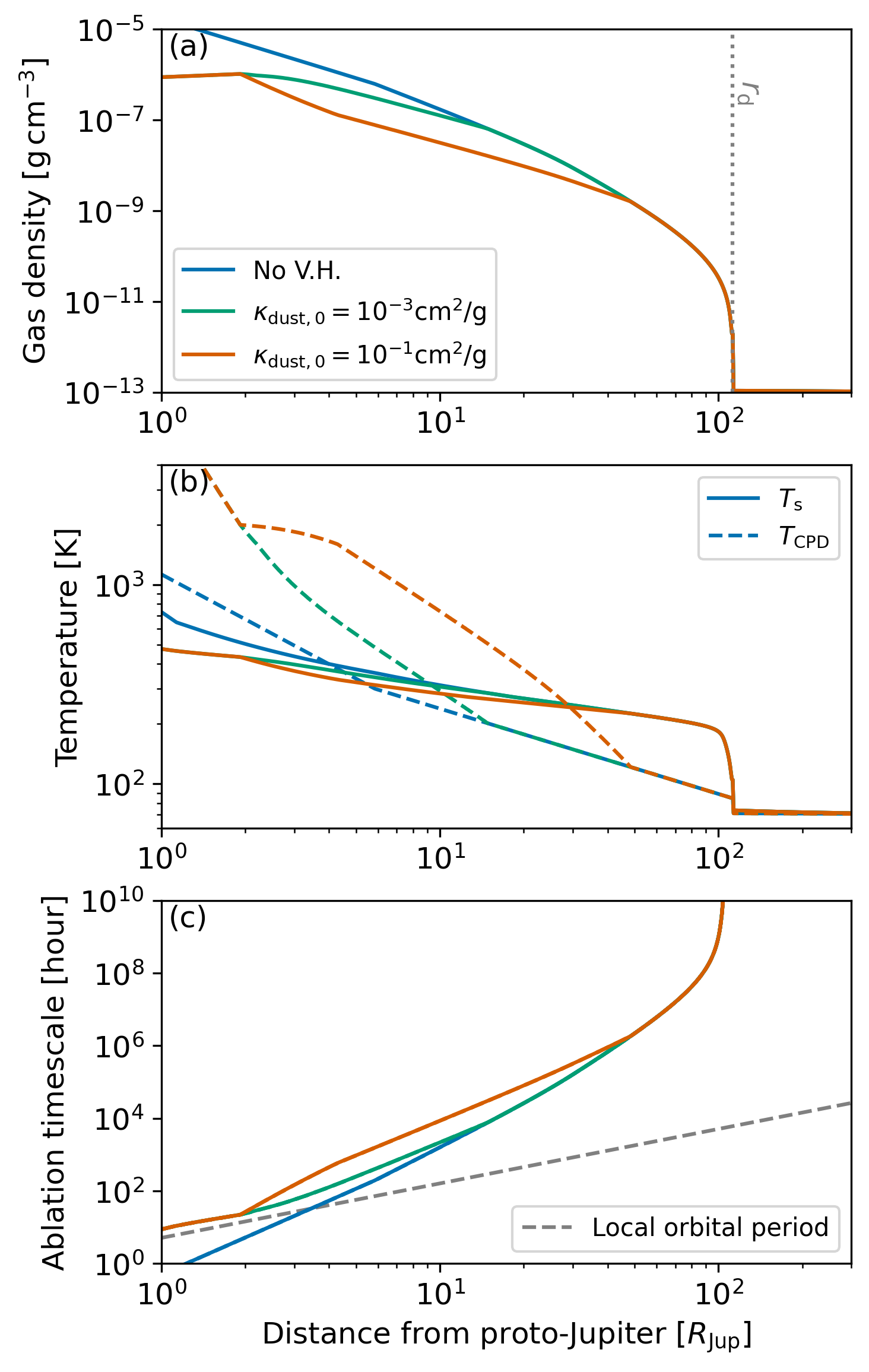}
    \caption{
    Radial profiles of the CPD structure and planetesimal-ablation conditions.
    {\bf (a)} Gas-density profiles in the CPD. The blue line shows the disk heated solely by irradiation, whereas the green and orange lines show disks including viscous heating with $\kappa_{\mathrm{dust},0}=10^{-3}~\mathrm{cm^2/g}$ and $\kappa_{\mathrm{dust},0}=10^{-1}~\mathrm{cm^2/g}$, respectively. The vertical dotted line marks the outer edge of the disk, $r_\mathrm{d}$.
    {\bf (b)} Temperature profiles in the CPD. The solid and dashed lines show the planetesimal surface temperature, $T_\mathrm{s}$, and the CPD midplane temperature, $T_\mathrm{CPD}$, respectively.
    {\bf (c)} Estimated ablation timescale. The dashed line shows the local orbital timescale.
    }
    \label{fig:disk_wVH}
\end{figure}

In the model including viscous heating, we determine the CPD temperature as
\begin{align}
T_\mathrm{CPD} = \max \left( T_\mathrm{thin}, T_\mathrm{flare}, T_\mathrm{vis} \right).
\end{align}
Because the surface density, temperature, and opacity are mutually coupled through Eqs.~\ref{eq:surface_density_CPD}, \ref{eq:T_CPD_vis}, and \ref{eq:kappa_disk_gas}, we solve these equations iteratively using the bisection method. Figrue~\ref{fig:disk_wVH} shows the CPD structure under the viscous-heating. We also plot the planetesimal's surface temperature and the ablation timescale, assuming a free-fall velocity for planetesimals. Table~\ref{tb:model_viscous_heating_disk} summarizes the parameters used in the viscously heated disk model.

\bibliography{sample701}{}
\bibliographystyle{elsarticle-harv}

\end{document}